\documentclass[11pt,twoside,letterpaper]{article} 
\usepackage{fancyhdr}
\usepackage[dvips]{graphicx}
\usepackage{rotating}

\usepackage{graphics}
\usepackage{graphicx}
\usepackage{amsmath}
\usepackage{amsfonts}
\usepackage{amssymb}
\usepackage{comment}
\usepackage{color}


\makeatletter
\def\cleardoublepage{\clearpage\if@twoside \ifodd\c@page\else%
    \hbox{}%
    \thispagestyle{empty}%
    \newpage%
    \if@twocolumn\hbox{}\newpage\fi\fi\fi}
\makeatother

\def\figurename{Figure}
\makeatletter
\renewcommand{\fnum@figure}[1]{\figurename~\thefigure.}
\makeatother

\def\tablename{Table}
\makeatletter
\renewcommand{\fnum@table}[1]{\tablename~\thetable.}
\makeatother

\usepackage{enumerate}
\usepackage{makeidx}
\usepackage[dvips]{graphicx}
\usepackage{color}
\usepackage{lineno}
\usepackage{epsf}
\usepackage{amsfonts}
\usepackage{amssymb}
\usepackage{amsmath}
\usepackage{amsthm}
\usepackage{palatino}
\usepackage{fancyhdr}

\newtheorem{theorem}{Theorem}

\newtheorem{definition}{Definition}

\title{Numerical and analytical methods for bond pricing in short rate convergence models of interest rates}
\author{Zuzana Bu\v{c}kov\'a \and  Be\'{a}ta Stehl\'{\i}kov\'{a} \and Daniel \v{S}ev\v{c}ovi\v{c}
\thanks{
Department of Applied Mathematics and Statistics, Comenius University, 842 48 Bratislava, Slovakia. Corresponding author: D. \v{S}ev\v{c}ovi\v{c}, sevcovic@fmph.uniba.sk
\newline
The research has been supported by VEGA 1/0251/16 project and FP7-PEOPLE-2012-ITN project \#304617 - STRIKE. 
\newline
This survey chapter has been submitted to the book collection: Interest Rates: Global Trends, Macroeconomic Implications and Analysis, 2016 Nova Science Publishers, Inc., Hauppauge. 
}
}

\date{}

\makeindex

\begin{document}

\maketitle

\begin{abstract}
In this survey paper we discuss recent advances on short interest rate models which can be formulated in terms of a stochastic differential equation for the instantaneous interest rate (also called short rate) or a system of such equations in case the short rate is assumed to depend also on other stochastic factors. Our focus is on convergence models, which explain the evolution of interest rate in connection with the adoption of Euro currency. Here, the domestic short rate depends on a stochastic European short rate. In short rate models, the bond prices, which determine the term structure of interest rate, are obtained as solutions to partial differential equations. Analytical solutions are available only in special cases; therefore we consider the question of obtaining their approximations. We use both analytical and numerical methods to get an approximate solution to the partial differential equation for bond prices.
\end{abstract}

\tableofcontents

\pagestyle{fancy}
\fancyhead{}
\fancyfoot{}
\fancyhead[EL,OR]{\thepage}
\renewcommand\headrulewidth{0.5pt}

\section{Introduction}
\label{sec:introduction}

\fancyhead[EC,OC]{Introduction}
\fancyhead[EL,OR]{\thepage}

An interest rate model is a description of interest rates' evolution\footnote{for example, the rate on one-year loan today and next year} and their dependence on maturity\footnote{interest rates on, for instance, one-year and ten-year loans are different}; the dependence of the interest rate on maturity is called the term structure of interest rates. Given the state of the market today, the future interest rates cannot be predicted exactly; the models gives their probability distribution. However, since the interest rates are interconnected, often only some underlying processes are modeled, which in turn determine the interest rates.

We deal with so-called short rate models, which are based on a theoretical quantity, the \textit{short rate}. It is a rate of interest for a default-free investment with infinitely small maturity. The other investments, with other maturities, include some risk: the evolution of the interest rates during the "life" of this investment can increase or decrease their value. Therefore it may not surprising that, besides the probabilistic description of the short rate evolution, there is another input - called market price of risk -  needed in order to compute the term structure of interest rates; cf. \cite[pp. 29-31]{fabozzi} for a further intuition following these ideas. 

Mathematical models can be described by solutions to linear parabolic differential equations, which degenerate to the hyperbolic ones at the boundary. Applying the Fichera theory to interest rates models one can treat the boundary conditions in a proper way. Correct treatment of boundary conditions is important in numerical schemes. 
 
We propose an approximate analytical solution for a class of one-factor models and derive the order of its accuracy. These models can be used to model the European short rate in convergence models. We show an example of a convergence model of this kind and the analytical approximation formula for domestic bond prices, together with the derivation of its accuracy. 

In some cases, a one-factor model is not sufficient to fit the European interest rates and we need a two-factor model to model the European short rate. Therefore, we also investigate a three-factor convergence model.

\section{Which model for term structures should one use?} 

\fancyhead[EC,OC]{Term structure models}
\fancyhead[EL,OR]{\thepage}

This is the title of the paper \cite{which}, in the beginning of which the author presents several criteria which a suitable model should have:

\begin{quote}
\textit{A practitioner wants a model which is
\begin{enumerate}[(a)]
\setlength{\itemsep}{0cm}%
\setlength{\parskip}{0cm}
\item{flexible enough to cover most situations arising in practice;}
\item{simple enough that one can compute answers in reasonable time;}
\item{well-specified, in that required inputs can be observed or estimated;}
\item{realistic, in that the model will not do silly things.}
\end{enumerate}
Additionally, the practitioner shares the view if an econometrician who wants
\begin{enumerate}[(e)]
\item{a good fit of the model to data;}
\end{enumerate}
and a theoretical economist would also require
\begin{enumerate}[(f)]
\item{an equilibrium derivation of the model.}
\end{enumerate}
}\end{quote}

Our work is mainly concerned with the point (b). Approximate analytical formulae enlarge the set of models for which \textit{one can compute answers in reasonable time}, as required above.
Moreover,  an easy computation of the observed quantities can significantly simplify a calibration of the model. Note that calibration of the model based on a comparison of market prices and theoretical prices given by the model often requires many evaluations of theoretical prices for different sets of parameters, as well as times to maturity and the short rate levels. 
Hence it is useful also to establish whether the point (e) above holds or not.

\section{Basic concepts of stochastic calculus }
 
\fancyhead[EC,OC]{Stochastic calculus}
\fancyhead[EL,OR]{\thepage}

In this section we briefly present the basic definitions and theorems of stochastic calculus which will be needed to formulate models considered here. For more details see, e.g., \cite{oksendal}, \cite{karatzas-shreve}. 

 \begin{definition} \cite[Definition 2.1.4]{oksendal}
A {stochastic process} is a parametrized collection of random variables  $\{ X_t \}_{t \in \mathcal{T}}$ defined on a probability space $(\Omega, \mathcal{F}, \mathcal{P})$ and assuming values in $\mathbb{R}^n$.
\end{definition}

An important stochastic process, used as a building block for other, more complicated processes, is a Wiener process.
\begin{definition} \cite[Definition 2.1]{sevcovic} \label{def:Wiener}
A stochastic process $\{w(t), t \geq 0\}$ is called a {Wiener process}, if it satisfies the following properties:
\begin{enumerate}[(i)]
\item{$w(0)=0$ with probability 1;}
\item{every increment $w(t+\Delta t)-w(t)$ has the normal distribution $N(0,\Delta t)$;}
\item{the increments $w(t_n)-w(t_{n-1})$, $w(t_{n-1})-w(t_{n-2})$, $\dots$, $w(t_2)-w(t_1)$ 
for $0 \leq t_1 < \dots < t_n$   are  independent.}
\end{enumerate}
\end{definition}
Existence of such a process can be asserted using the Kolmogorov extension theorem, which builds a stochastic process from its finite dimensional distributions, cf. \cite[Chapters 2.1 and 2.2]{oksendal}, \cite[Chapter 2.2]{karatzas-shreve}.

Using a Wiener process, we are able to define new processes. It would be useful to be able to use some kind of "noise" in the ordinary differential equations and a Wiener process provides a way of doing so. This leads to so called stochastic integrals and stochastic differential equations. Again, we follow the main ideas of \cite{oksendal}.

The first idea might be to consider an equation of the form
\begin{equation}
\frac{d X}{d t} = b(t,X_t)   + \sigma(X_t,t) \, u_t, \label{ito1}
\end{equation}
where the  term $u$ denotes some "noise", which should be stationary, with values at different time  being independent and having a zero expected value. However, there is no continuous process satisfying these conditions. Moreover, as a function on $[0,\infty) \times \Omega$ it cannot be even  measurable (considering Borel-measurable sets on $[0,\infty)$), see \cite[pp. 21-22]{oksendal} and references therein. Therefore, another approach is used. We write (\ref{ito1}) in a discrete form as
$$X_{t_{k+1}} = X_{t_k} + b(t,X_t) (t_{k+1} - t_k) + \sigma(X_t,t) \, u_{t_k} (t_{k+1} - t_k),$$
where $0=t_0 < t_1 < \dots < t_m=t$ is a partition of the interval $[0,t]$.
Recalling the desirable properties of the noise, the term $ u_{t_k} (t_{k+1} - t_k)$ should have stationary independent increments, which suggests using a Wiener process $w_{t_k}$. Then we have an equation 
$$X_{t_{k+1}} = X_{0} + \sum_{j=0}^{k-1} b(t,X_{t_j}) (t_{j+1} - t_j) + \sum_{j=0}^{k-1} \sigma(X_{t_k},t) 
\,(w_{t_{k+1}} - w_{t_{k}})$$
and if we are able to make a limit of the last sum in some "reasonable way", by denoting it by $\int_0^t \sigma(s,X_s) \, d w_s$ we can write
\begin{equation}
X_t = X_0 + \int_0^t b(s,X_s) \, d s + \int_0^t \sigma(s,X_s) \, d w_s. \label{ito2}
\end{equation}
This can indeed be done; in several ways, in fact, which leads to different kinds of stochastic integrals (It\=o vs. Stratonovich). We use {It\=o integral}, see the cited references \cite{oksendal} for details on its construction.

Finally, let us note that the equation (\ref{ito2}) is often written in a "differential form"
 \begin{equation}
d X_t = b(t,X_t) \, d t + \sigma(t,X_t) \, d w_t \label{eq:sde}
\end{equation}
which is called a {stochastic differential equation}. 

The computation of the "differential" $d Y_t$, where $Y_t$ is defined as $Y_t=f(t,X_t)$, where $f$ is a smooth function and $X$ satisfies the stochastic differential equation (\ref{eq:sde}) is performed via a stochastic generalization of the chain rule known from calculus. This can be done precisely using the integral representation of the stochastic processes (cf. \cite[pp. 150-153]{karatzas-shreve}) and results in the famous {It\=o lemma}. 
We provide its formulation for the case of a one-dimensional process from \cite{oksendal}.

\begin{theorem} \cite[Theorem 4.1.6]{oksendal}
Let $X_t$ be an It\=o process given by $$d X_t = u(t, X_t) \: d t + v(t, X_t)  \: d w.$$ Let $g(t,x) \in C^2([0,\infty) \times \mathbb{R})$. Then $Y_t=f(t,X_t)$  is again an It\=o process and
$$d Y_t = \frac{\partial g}{\partial t}(t,X_t) d t + \frac{\partial g}{\partial x}(t, X_t) d X_t + 
\frac{1}{2} \frac{\partial^2 g}{\partial x^2}(t,X_t) (d X_t)^2,$$
where $(d X_t)^2 = (d X_t)(d X_t)$ is computed according to the "rules"
$$ d t\, d t = d t \, d w_t = d w_t \, d t=0,  d w_t \, d w_t = d t.$$
\end{theorem}
 
A multidimensional formulation can be found for example in \cite[Theorem 4.2.1]{oksendal}, \cite[Theorem 3.6]{karatzas-shreve} or in the original paper by Kiyoshi It\=o 
\cite[Theorem 6]{ito}.

\begin{figure}
\centerline{
\includegraphics[width=0.5\textwidth]{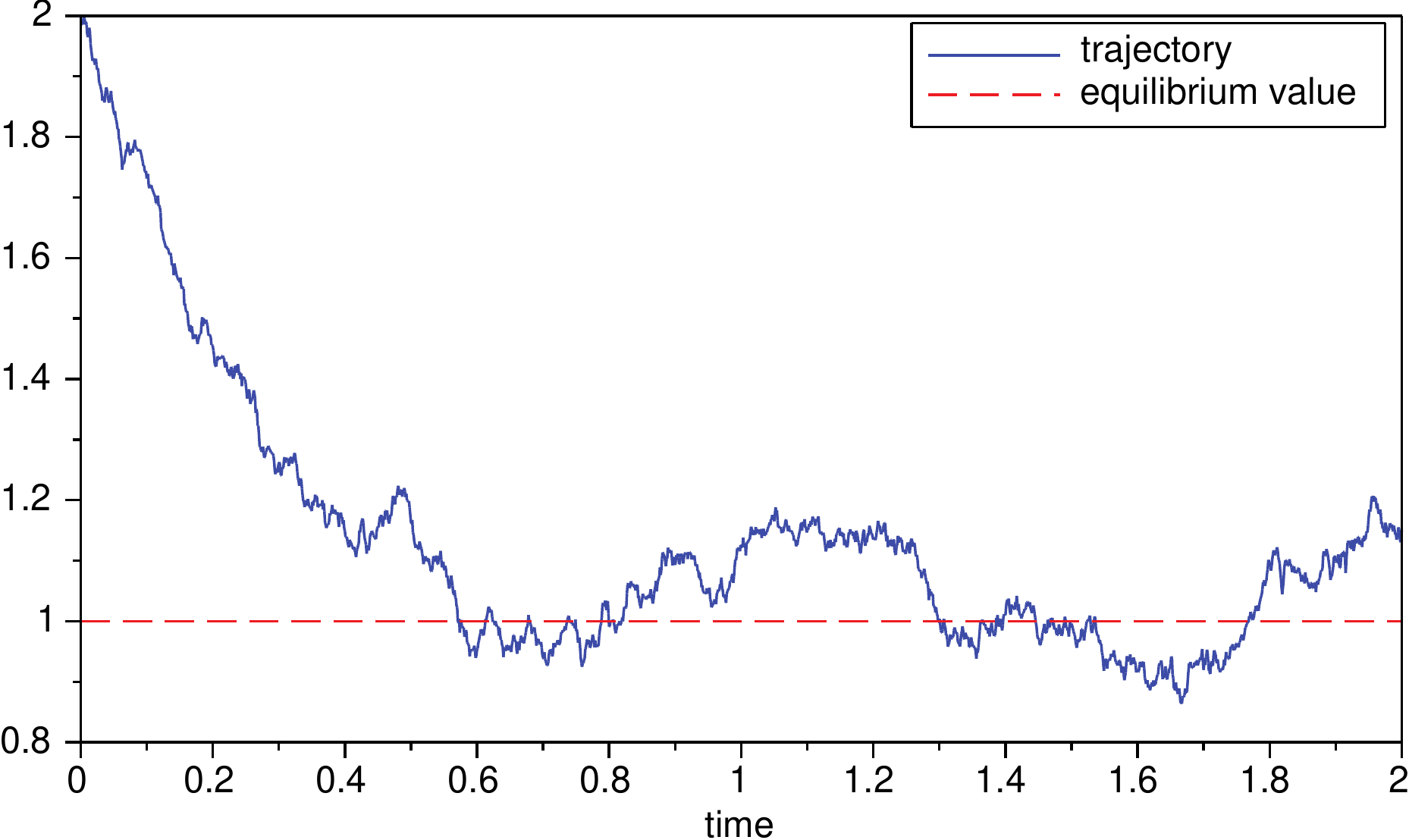}
}
\caption{Sample path of an Ornstein-Uhlenbeck process.}
\label{fig:ou}
\end{figure}

In order to illustrate It\=o's process, we present an example of a stochastic differential equation which will be useful later. It  describes the evolution of a so called {Ornstein-Uhlenbeck process}:
\begin{equation}
d x = \kappa (\theta - x) \, d t + \sigma \, d w,
\end{equation}
where $\kappa, \theta$ and $\sigma$ are positive constants. Without the stochastic $d w$ term, it would be an ordinary differential equation with the solution $x_t = x_0 e^{-\kappa t} + \theta (1 - e^{-\kappa t}),$ where $x_0$ is the value of the process at time $t=0$. With the stochastic term included, the solution becomes a random variable and can be written in an explicit form
$$x_t = x_0 e^{-\kappa t} + \theta (1 - e^{-\kappa t}) + \sigma \int_0^t d w.$$
The trend, reversion to the equilibrium level $\theta$, whose speed depends on $\kappa$, is preserved (processes with this property are called {mean-reversion processes}). Furthermore, there are random fluctuations around this trend; their impact depends on the parameter $\sigma$. Sample trajectory of an Ornstein-Uhlenbeck process is presented in Figure \ref{fig:ou}.

Similarly as in the case of ordinary differential equations, a closed-form solution is not always available, but  numerical approximations are still possible. The simplest one is the {Euler-Maruyama} scheme, which is a generalization of the Euler method known from numerical methods for ordinary differential equations. It consists of replacing the differentials in (\ref{eq:sde}) by finite differences and simulating the increments of a Wiener process:
\begin{eqnarray}
X_0 &=& x_0, \nonumber \\
X_{t+\Delta t}  &=& X_t + b(t,X_t) \, \Delta t   + \sigma(t,X_t) \, \Delta w_t, \nonumber
\end{eqnarray}
where $\Delta w$ are independent realizations from $\mathcal{N}(0,\Delta t)$ distribution. There are also other methods, which have a higher precision (for example, Milstein scheme, Runge-Kutta methods, cf. \cite{seydel} for an introduction or \cite{kloeden-platen} for more details).

\section{Short rate models  }

\fancyhead[EC,OC]{Short rate models}
\fancyhead[EL,OR]{\thepage}

\begin{figure}
\centerline{
\includegraphics[width=0.5\textwidth]{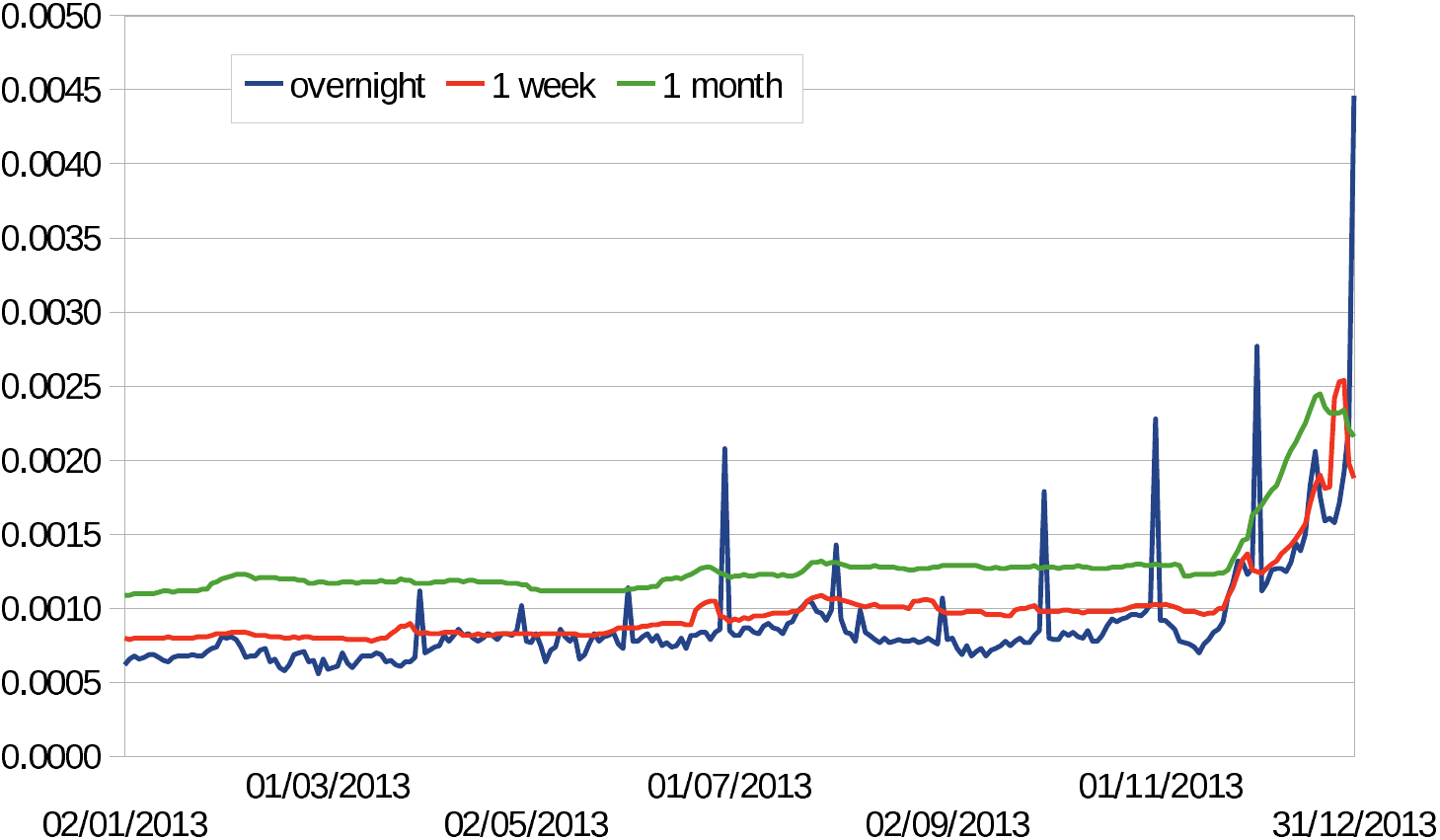}
}
\caption{Euro interest rates with short maturities  - possible approximations of short rate. Data source: \textit{http://www.emmi-benchmarks.eu}}
\label{fig:short-rate-example}
\end{figure}

Short rate models are formulated in terms of a stochastic differential equation (one-factor models) a system of stochastic differential equations (multi-factor models) determining the short rate, see Figure \ref{fig:short-rate-example} for an example of market data which - being interest rates with short maturities - can be thought of as approximations of the theoretical short rate. 

We start with a simple stochastic differential equation which describes some popular features of the market rates. Then, seeing the shortcomings of the models, we move to more complicated ones. Each of them addresses a specific feature and the choice of the model needs to take this into account. For selected stochastic processes we explain the motivation that leads to considering them  as a model for the short rate.

\begin{figure}[th]
\centerline{
\includegraphics[width=0.5\textwidth]{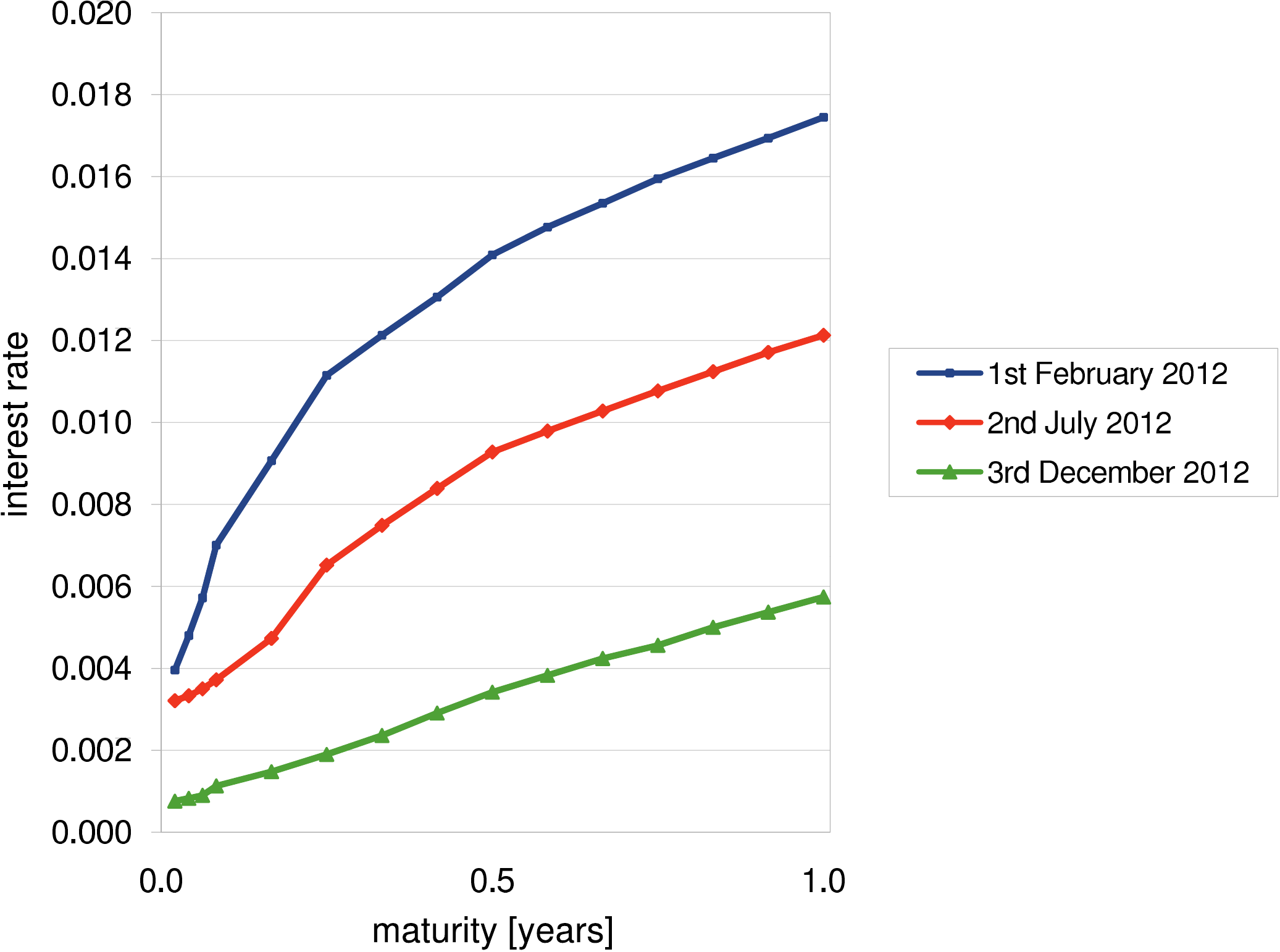}
}
\caption{Euro interest rates - examples of term structures. Data source: \textit{http://www.emmi-benchmarks.eu}}
\label{fig:term-structure-example}
\end{figure}

We also discuss bond prices. A zero-coupon bond is a financial security that pays a unit amount money to its holder at the specified time of maturity. The bond prices $P=P(t,T,\mathbf{x})$ (where $t$ is time, $T$ is time to maturity and $\mathbf{x}$ is a vector of factors determining the short rate) are then connected with interest rates $R=R(t,T,\mathbf{x})$ through the formula
\begin{equation}
P(t,T,\mathbf{x}) = e^{-R(t,T,\mathbf{x})(T-t)}, \; \textrm{ i.e., } \; R(t,T,\mathbf{x}) = - 
\frac{ \textrm{ln}\,P(t,T,\mathbf{x})}{T-t}.
\end{equation}
Examples of interest rates with different maturities can be seen in Figure \ref{fig:term-structure-example}.

In short rate models, the prices of bonds (as well as other interest rate derivatives) are solutions to a parabolic partial differential equation. Even in a case of a derivative with such a simple payoff, as it is the case of a bond, closed form solution is available only in very special cases. The later topics presented in this paper are then connected by a pursuit of finding approximations of the bond prices (and hence also term structures) in those cases when they are not known in a closed form.

\subsection{One-factor models}

When speaking of one-factor short rate models, the term \textit{one-factor} refers to the fact that there is \textit{one} Wiener process used in the definition of the short rate process, i.e., there is \textit{one} source of randomness. 

Hence, there is a scalar stochastic differential equation for the short rate $r$, which can be written in a general form as 
$$d r = \mu(r,t) d t + \sigma(r,t) d w,$$
where $w$ is a Wiener process. Recall from the section on stochastic processes that the function $\mu(r,t)$ determines the trend of the process, while the function $\sigma(r,t)$ determines the nature of the random fluctuations. Specifying the functions $\mu(r,t)$ and $\sigma(r,t)$ characterizes the short rate model.

Let $P=P(r,t)$ be the price of a derivative at time $t$ when the current level of the short rate is $r$, which pays a given payoff at time $T$. We consider a construction of a portfolio consisting of derivatives with two different maturities, continuously rebalanced so that the risk coming from the Wiener process is eliminated\footnote{It can be shown that it is possible if we assume an "idealized market" with no transaction costs, ability to buy or sell any desired amount of a security for its present price, to borrow/lend any amount of money for the short rate interest rate and operating in continuous time. 
 This idealization of reality in the derivation of the equation for security prices might be another reason for being "satisfied" with a meaningful simple approximation of the short rate process, instead of requiring an extremely complex model for it.}. Then, to eliminate a possibility of an arbitrage, the return of such a portfolio has to be equal to the current short rate, which leads to a partial differential equation for the derivative price $P$, which reads as
$$\partial_t P + (\mu(r,t) - \lambda(r,t) \sigma(r,t)) \partial_r P + \frac{1}{2} \sigma(r,t)^2 \partial^2_{r} P = 0$$
for all admissible values of $r$ and for all $t \in [0,T)$. We refer to \cite{kwok}, \cite{sevcovic}, for more details on the derivation of the partial differential equation. Here and after we denote by $\partial_t P, \partial_r P$ the first partial derivatives of $P$ with respect to $t$, $r$ and the second derivative $\partial^2_r P$ of $P$ with respect to $r$. Note that the equation includes a new function $\lambda(r,t)$. It appears during the derivation of the equation, when it turns out that a certain quantity, measuring the rise of the expected return for one unit of risk, has to be independent of the maturity $T$. It is denoted by $\lambda(r,t)$ and because of its interpretation it is called market price of risk.  It is necessary to include it into the specification of a model when we want to price derivatives, in addition to talking about the short rate evolution. Note that the equation holds for any derivative, the specific derivative determines the terminal condition $P(r,T)$ which equals the security payoff.

If we consider only Markov models, i.e.,  $\mu$, $\sigma$ and $\lambda$ are functions only of the variable $r$ and do not explicitly depend on time $t$ (which will be the case for the models studied in this thesis), it is convenient to introduce a new variable $\tau=T-t$ denoting time remaining to maturity. For the bond price we obtain the partial differential equation (PDE) 
\begin{eqnarray}
-\partial_{\tau} P + (\mu(r) - \lambda(r) \sigma(r)) \partial_r P + \frac{1}{2} \sigma(r)^2 \partial^2_{r} P &=& 0\,\,\,  \textrm{ for all } r \textrm{ and } \tau \in (0,T], \label{eq:pde-bond-1f} \\
P(r,0)&=&1 \,\,\,  \textrm{ for all } r. \label{eq:pde-bond-1f-}
\end{eqnarray}

Alternatively, a model can be formulated in the so-called risk-neutral measure $\mathbb{Q}$, which is an equivalent probability measure to $\mathbb{P}$, in which the process is physically observed. In the risk-neutral measure, the prices of the securities can be expressed in a form of expected values. The change of measure is related to the market price of risk from the partial derivative approach above and mathematically it is based on Girsanov theorem (cf. \cite[Section 8.6]{oksendal}). The general model above in the risk neutral model reads as
\begin{equation}
\mathrm{d}r= \tilde{\mu}(r) \mathrm{d}t + \tilde{\sigma}(r) \mathrm{d}w^{\mathbb{Q}},
\end{equation}
where $w^{\mathbb{Q}}$ is a Wiener process under the risk neutral measure, while the risk-neutral drift and volatility are given by
\begin{equation}
\tilde{\mu}(r)  = \mu(r) - \lambda(r) \sigma(r), \tilde{\sigma}(r)  = {\sigma}(r),
\end{equation}
cf. \cite[Section 7.2]{kwok2}. Comparing this with (\ref{eq:pde-bond-1f}) we can see that the risk-neutral formulation contains all information needed to write the valuation PDE. Therefore, when dealing with pricing bonds or other derivatives, the model is often formulated in the risk-neutral form. Finally, let us note that the two alternative expressions for the   prices - expected values under the risk-neutral measure and solutions to  partial differential equations - are related also via so-called Feynman-Kac formula, cf. \cite[Theorem 8.2.1]{oksendal}.

\subsection{Vasicek and Cox-Ingersoll-Ross models}

Recall that the Ornstein-Uhlenbeck process is a stochastic process given by
\[
d r= \kappa(\theta -r) \: d t + \sigma \: d w,
\]
where $\kappa, \theta, \sigma >0$ are given constants and $w$ is a Wiener process. This process can be used as a simple model for the short rate, known as Vasicek model, as it has been suggested in \cite{vasicek} by Old\v rich Va\v s\'i\v cek. He defined the market price of risk to be equal to a constant $\lambda$, which results in the partial differential equation for the bond prices that reads as  (recall  its general form (\ref{eq:pde-bond-1f})-(\ref{eq:pde-bond-1f-}))
\begin{equation}
-\partial_{\tau} P + (\kappa(\theta-r)- \lambda \sigma ) \partial_r P + \frac{1}{2} \sigma^2  \partial^2_{r} P = 0 \label{eq:pdr:vas}
\end{equation}
for all $r$  and  $\tau \in (0,T]$, and 
$P(r,0)=1$ for all $r$. 
This differential equation can be solved explicitly; its solution has the form 
\begin{equation}
P(r,\tau)=A(\tau)e^{-B(\tau)r} \label{eq:vasicek-form}
\end{equation}
 and the functions $A,B$ are given by (see \cite{vasicek})
\begin{equation}
\ln A(\tau) = \left( -\theta + \frac{\lambda \sigma}{\kappa} + \frac{\sigma^2}{2 \kappa^2} \right)
\left(-\frac{1-e^{-\kappa \tau}}{\kappa}+\tau \right)
 -\frac{\sigma^2}{4 \kappa^3} (1-e^{-\kappa \tau})^2, \; B(\tau) = \frac{1-e^{-\kappa \tau}}{\kappa}.
\label{vasicek-bond}
\end{equation}

One of the consequences of the  constant volatility is a conditional normal distribution of the future interest rates and thus a possibility of negative interest rates. Historically, this was been one of the motivations for proposing other short rate models. Note, however, that while some of the interest rates observed in these days can be indeed negative, the negative values of the Ornstein-Uhlenbeck stochastic process is not consistent with absence of arbitrage in the context of default intensity models \cite{kane}  which leads to solving exactly the same parabolic PDEs. A popular alternative is the Cox-Ingersoll-Ross model \cite{cir} (usually abbreviated as CIR model) which does not allow negative interest rates, while it preserves analytical tractability of bond prices. The stochastic differential equation for the  short rate is given by 
\begin{equation}
d r= \kappa(\theta -r) \: d t + \sigma \sqrt{r} \: d w, \label{eq:cir_sde}
\end{equation}
with $\kappa, \theta, \sigma >0$ being constants. The difference from the Vasicek model lies in the volatility, which is now equal to $\sigma \sqrt{r}$. Intuitively, if the short rate $r$ is small, then also the volatility is small; if short rate hits zero, the volatility becomes zero as well and the positive drift pushes the short rate to a positive value. It can be shown 
that the process is indeed nonnegative for all times and,  moreover, if the condition $2 \kappa \theta > \sigma^2$ is satisfied, the process remains strictly positive. If the market price of risk is chosen to be $\lambda \sqrt{r}$, the equation (\ref{eq:pde-bond-1f}) with initial condition (\ref{eq:pde-bond-1f-}) becomes 
\begin{equation}
-\partial_{\tau} P + (\kappa(\theta-r)- \lambda \sigma r) \partial_r P + \frac{1}{2} \sigma^2 r \partial^2_{r} P = 0 \label{eq:pdr:cir}
\end{equation}
for all $r$  and  $\tau \in (0,T]$, and 
$P(r,0)=1$ for all $r$. 
Again, it can be solved in a closed form, assuming the solution (\ref{eq:vasicek-form}), inserting it into the partial differential equation and obtaining a system of ordinary differential equations for the functions $A(\tau), B(\tau)$. This system can be solved explicitly, see \cite{cir} for the exact formulae. 

\subsection{Chan-Karolyi-Longstaff-Sanders short rate model}

As we have seen, changing the constant volatility from the Vasicek model to $\sigma \sqrt{r}$ in CIR model prevents the short rate from becoming negative. However, the same reasoning applies to any volatility of the form $\sigma r^{\gamma}$ with $\gamma >0$.
Models with general $\gamma$ may perform better when applied to real data and the hypothesis of $\gamma=1/2$ is actually often rejected by statistical tests. 

The pioneering paper \cite{ckls} by Chan, Karolyi, Longstaff and Sanders started the discussion on the correct form of the volatility. Authors used  proxy for the short rate process and 
considered a general short rate model expressed in terms of a single stochastic differential equation 
\begin{equation}
d r=(\alpha+\beta r)\,d t+\sigma r^{\gamma} \,d w, \label{eq-ckls-}
\end{equation}
which has become known as the CKLS model. It includes Vasicek ($\gamma=0$) and CIR ($\gamma=1/2$) models as special cases (and thus allows testing them as statistical hypotheses on the model parameters), as well as several other models, see Table \ref{table:ckls-1}.
Chan \textit{et al.} estimated the parameters using the generalized method of moments. They found the parameter $\gamma$ to be significantly different from the values indicated by Vasicek and CIR models, see Table \ref{table:ckls-2}.

\begin{table}
\caption{One-factor short rate models considered in  \cite{ckls} as special cases of the stochastic process (\ref{eq-ckls-}).}
\begin{center}
\small
\begin{tabular}{l|l}
Model: & Equation for the short rate: \\ 
\hline
 Merton  \cite{irmodels-merton}  & $dr=\alpha dt + \sigma dw$ \\ 
 Vasicek \cite{vasicek} & $dr=(\alpha+\beta r)dt+\sigma dw$ \\ 
 Cox-Ingersoll-Ross (1985) \cite{cir} & $dr=(\alpha+\beta r)dt+\sigma r^{1/2} dw$ \\ 
 Dothan \cite{irmodels-dothan}, \cite{irmodels-br-sch-77} & $dr=\sigma r dw$ \\ 
 Geometrical Brownian motion \cite{irmodels-marsh} & $dr=\beta r dt + \sigma r dw$ \\ 
 Brennan-Schwartz \cite{irmodels-br-sch-80}, \cite{irmodels-crank-nicolson-numerics} & $dr=(\alpha + \beta r) dt + \sigma r dw$ \\ 
 Cox-Ingersoll-Ross (1980) \cite{irmodels-cir-80}  & $dr=\sigma r^{3/2} dw$ \\ 
 Constant elasticity of variance \cite{irmodels-marsh} & $dr = \beta r dt + \sigma r{^\gamma} dw$ \\ 
\hline 
\end{tabular} 
\end{center}
\label{table:ckls-1}
\end{table}

\begin{table}
\caption{Parameters estimates and results from testing the hypotheses given by Vasicek and CIR models in \cite{ckls}.}
\begin{center}
\small
\begin{tabular}{l|cccc|c}
Model: & $\alpha$  & $\beta$ & $\sigma^2$ & $\gamma$ &  P-value \\ 
\hline
unrestricted & 0.0408 & -0.5921 & 1.6704 & 1.4999 & - \\
  Vasicek & 0.0154 & -0.1779 & 0.0004 & 0 & 0.0029 \\
CIR & 0.0189 & -0.2339 & 0.0073 & 1/2 & 0.0131  \\
\hline 
\end{tabular} 
\end{center}
\label{table:ckls-2}
\end{table}

A modification of the generalized method
of moments (so called robust generalized method of moments), which is robust to a
presence of outliers, was developed in \cite{aquilla}.
Another contribution to this class of estimators is for example 
indirect robust estimation by \cite{czellar}.
Another popular method for parameter estimation are Nowman's Gaussian estimates
\cite{nowman}, based on approximating the likelihood function. They were used in \cite{episcopos}
for a wide range of interest rate markets. 
There are several other calibration methods for the short rate process, such as quasi maximum
likelihood, maximum likelihood based on series expansion of likelihood function
by A\"it-Sahalia \cite{ait-sahalia-transition}, Bayesian methods such as Markov chain Monte Carlo and others.

A common feature of these approaches is taking a certain market rate as a proxy to the short rate 
and using the econometric techniques of time series analysis to estimate the parameters of the model. These parameters can be 
 used afterwards to price the bonds and other derivatives. For example in  \cite{nowman-2}, the parameters of the CKLS process were first estimated using the Nowman's methodology and afterwards derivatives prices were computed by numerically solving the partial differential equation using the Box method. For more results of this kind see \cite{nowman-3}, \cite{nowman-4}.

An alternative would be using the derivatives prices to calibrate the parameters of the model. This, however, requires a quick computation of the prices, since they have to be computed many times with different parameters during the calibration procedure. Exact solution to the bond pricing equation available for Vasicek and CIR model made this possible in the case of these two models, 
cf.~\cite{sevcovic-csajkova-2}, \cite{sevcovic-csajkova-1}. In general, when 
the exact solution is not available, approximate analytical solution provides a convenient alternative.

\subsection{Other one-factor models}

Modifying the constant volatility is not the only way for ensuring  positivity of short rate. Another simple way is defining short rate as a positive function whose argument is a stochastic process. In particular, Black-Karasinski model \cite{black-karasinski} (also called exponential Vasicek because of its construction, cf. \cite[Section 3.2.5]{brigo-mercurio}) defined the short rate as $r=e^x$, where $x$ follows an Ornstein-Uhlenbeck process
\begin{equation}
d x = \kappa(\theta-x)\,   d t + \sigma \,  d w. \label{eq:bk-1}
\end{equation}

Note that in the case of Black-Karasinski model, the stochastic differential equation for the short rate $r$ reads as
$$d r= r ( \kappa \theta + \frac{1}{2} \sigma^2 - \kappa \textrm{ ln}\,r) \, d t + \sigma r \,  d w,$$
which means that the short rate does not have a linear drift, common to the previously considered models. 

\begin{figure}
\centerline{
\includegraphics[width=0.6\textwidth]{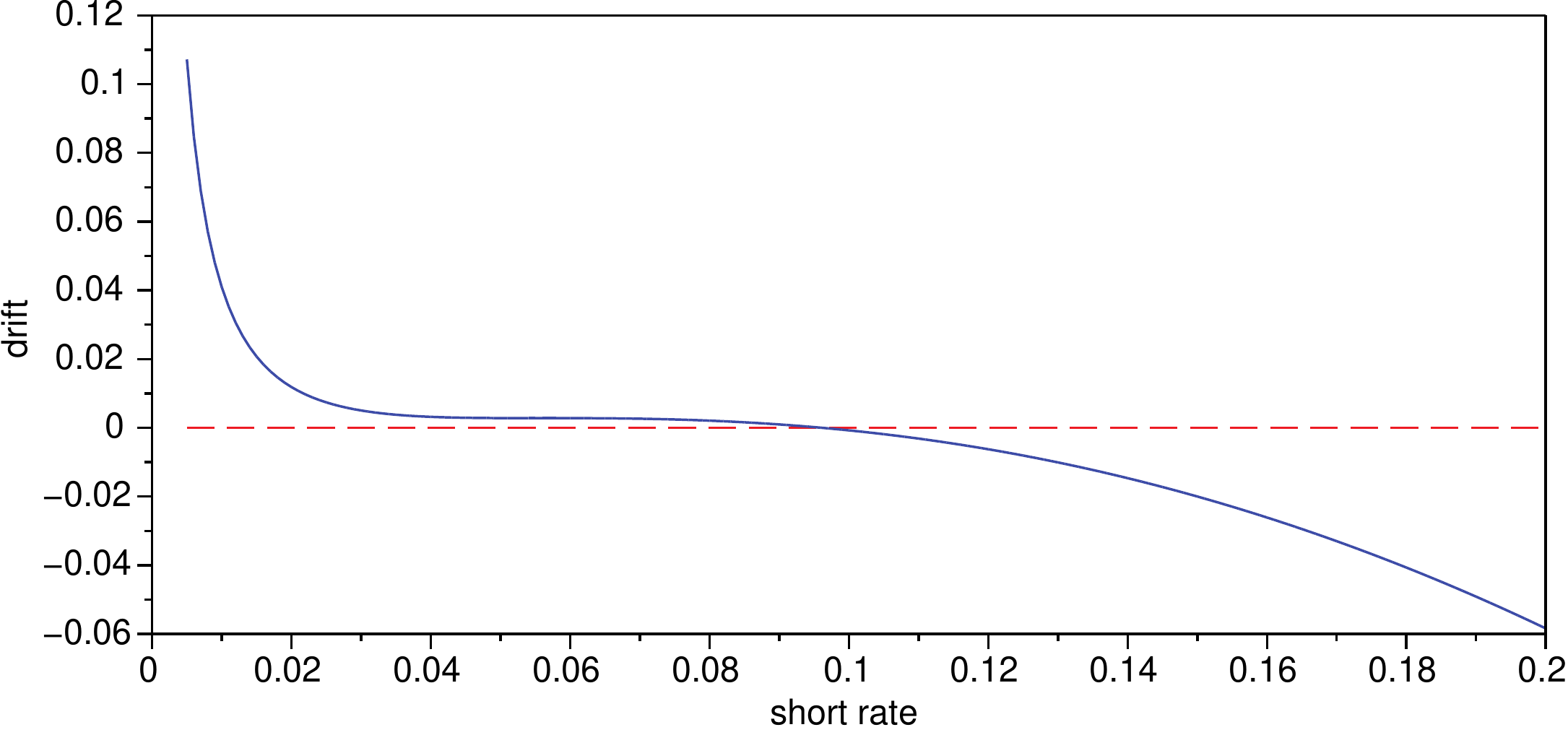}
}
\caption{Nonlinear drift of the A\"it-Sahalia model  \cite{ait-sahalia-drift} for parameters $\alpha_{-1} = 0.000693, \alpha_0 = -0.0347, \alpha_1 = 0.676, \alpha_2 = -4.059$, taken from \cite{ait-sahalia-transition}}
\label{fig:ait-sahalia-drift}
\end{figure}

Another nonlinear-drift model has been suggested by A\"it-Sahalia in \cite{ait-sahalia-drift} to produce very little mean reversion while the interest rates remain in the middle part of their domain, and strong nonlinear mean reversion at either end of the domain. This property is achieved
by the stochastic differential equation
$$d r = (\alpha_{-1} r^{-1} + \alpha_0 + \alpha_1 r + \alpha_2 r^2) \: d t + \sigma r^{\gamma} d w,$$
see Figure \ref{fig:ait-sahalia-drift} for a plot of the drift function for $\alpha_{-1} = 0.000693, \alpha_0 = -0.0347, \alpha_1 = 0.676, \alpha_2 = -4.059$ which are taken from \cite{ait-sahalia-transition}.

\subsection{Short rate as a sum of multiple factors}

\begin{figure}
\centerline{
\includegraphics[width=0.5\textwidth]{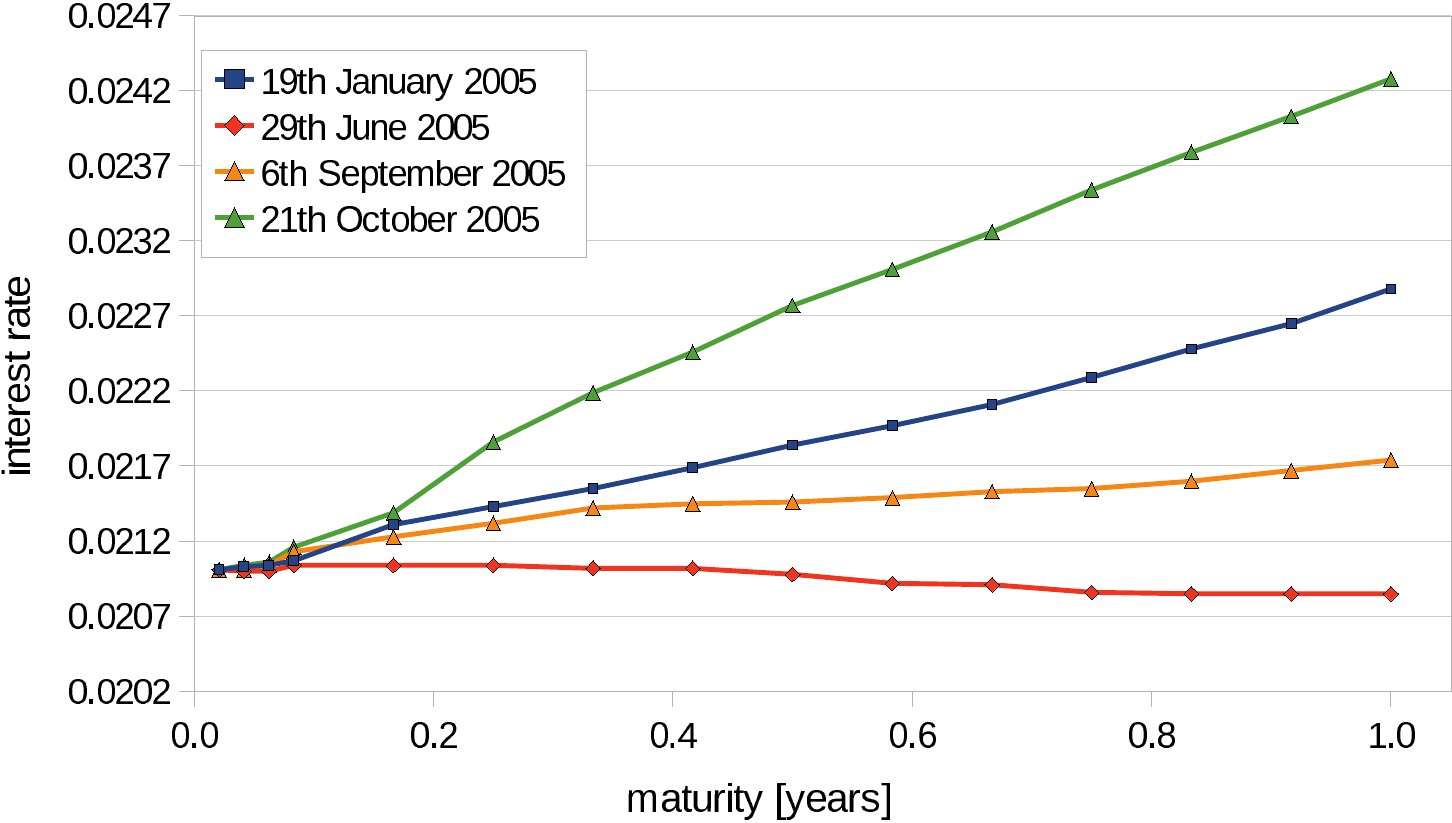}
}
\caption{Euro interest rates - examples of term structure starting from the same point. Data source: \textit{http://www.emmi-benchmarks.eu}}
\label{fig:multifactor-motivation}
\end{figure}

One of the consequences of using a one-factor short rate model is the bond price of the form $P=P(\tau,r)$. This means that the bond price with a given maturity is uniquely determined by the short rate level. Translating this into the language of term structures: the term structure is uniquely determined by its beginning (interest rate for infinitesimally small maturity, i.e., the short rate). While this might not be an unreasonable property of the interest rates in certain time periods, it clearly does not hold in others, as demonstrated in Figure \ref{fig:multifactor-motivation}.

If we define the short rate as a function of more factors, i.e., $r=r(x_1,\dots,x_n)$, then the bond price has the form $P=P(\tau,x_1,\dots,x_n)$. If the same short rate level can be achieved for several combinations of the factors $x_1,\dots,x_n$, these can produce different bond prices and, consequently, term structures - such as those seen in Figure \ref{fig:multifactor-motivation}.
Moreover, the factors determining the short rate may have a plausible interpretation on their own.

In \cite{babbs-nowman} the authors propose the model for the short rate $r$ to be
$$r = \mu - \sum_{j=1}^n x_i,$$
where $\mu$ is interpreted as the long-run average rate and
$x_1, \dots,, x_n$ represent the current effect of $n$ 
streams of economic "news", among which they include rumors about central bank decisions, economic statistics, etc.  The arrival of each of these news is modeled by the process
$$d x_i = \xi_i x_i \, d t + \sigma_i \, d w_i$$
with negative constants $\xi_i$ and possibly correlated Wiener processes $w_i$. Thus, the impact of any news dies away exponentially.
If the market prices of risk are taken to be constant, it is possible to express the bond prices in a closed form.

\begin{figure}
\centerline{
\includegraphics[width=0.5\textwidth]{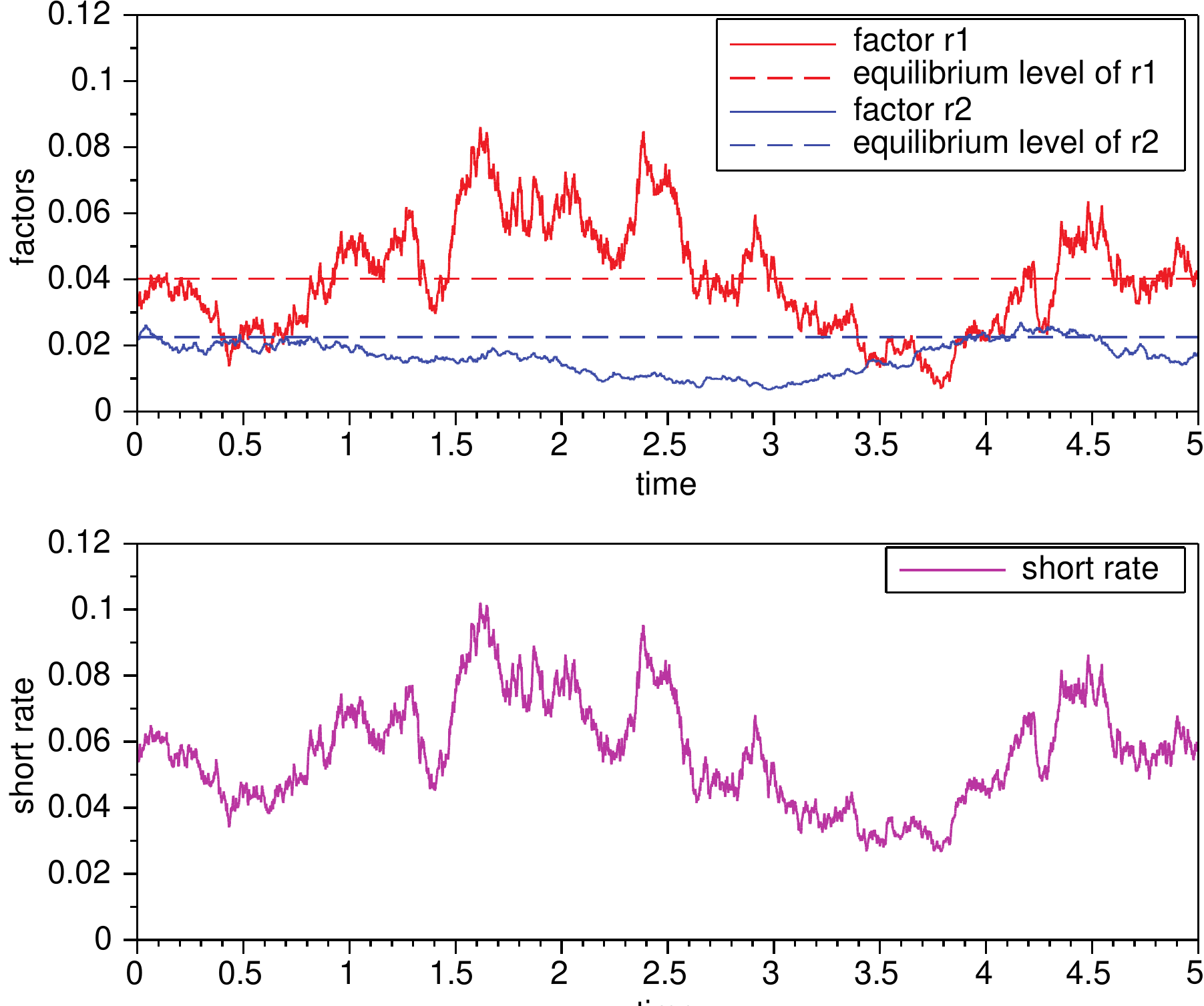}
}
\caption{Two-factor CIR model: sample trajectories of the factors and the short rate for parameters $\kappa_1=0.7298, \theta_1=0.04013, \sigma_1=0.16885$, 
$\kappa_2=0.021185, \theta_2=0.022543, \sigma_2=0.054415$,   taken from \cite{chen-scott}. }
\label{fig:2fCIR}
\end{figure}

A multi-factor version of a one-factor CIR model is formulated in \cite{chen-scott}, where the short rate $r$ is a sum of $n$ components, i.e., 
\begin{equation}
r=\sum_{j=1}^n r_i, \label{eq:n-CIR-1}
\end{equation}
with each $x_i$ following a Bessel square root process
\begin{equation}
d r_i = \kappa(\theta - r_i) \, d t + \sigma_i \sqrt{r_i} \, d w_i, \label{eq:n-CIR-2}
\end{equation}
assuming independent Wiener processes. Their independence and the choice of market prices of risk to be $\lambda_i \sqrt{r_i}$ again allows analytical expressions for the prices of bonds. In Figure \ref{fig:2fCIR} we show sample trajectories of a two-factor CIR model with parameters equal to $\kappa_1=0.7298, \theta_1=0.04013, \sigma_1=0.16885$, 
$\kappa_2=0.021185, \theta_2=0.022543, \sigma_2=0.054415$, which are taken from \cite{chen-scott}.

The equations (\ref{eq:n-CIR-1})-(\ref{eq:n-CIR-2}) can be generalized to general CKLS processes (\ref{eq-ckls-}) and correlated Wiener processes. However, with the exception of the special cases above, the closed form formulae for bond prices are not available and, therefore, their approximations are necessary.

\subsection{Stochastic volatility multiple-factor interest rate models}

\begin{figure}
\centerline{
\includegraphics[width=0.6\textwidth]{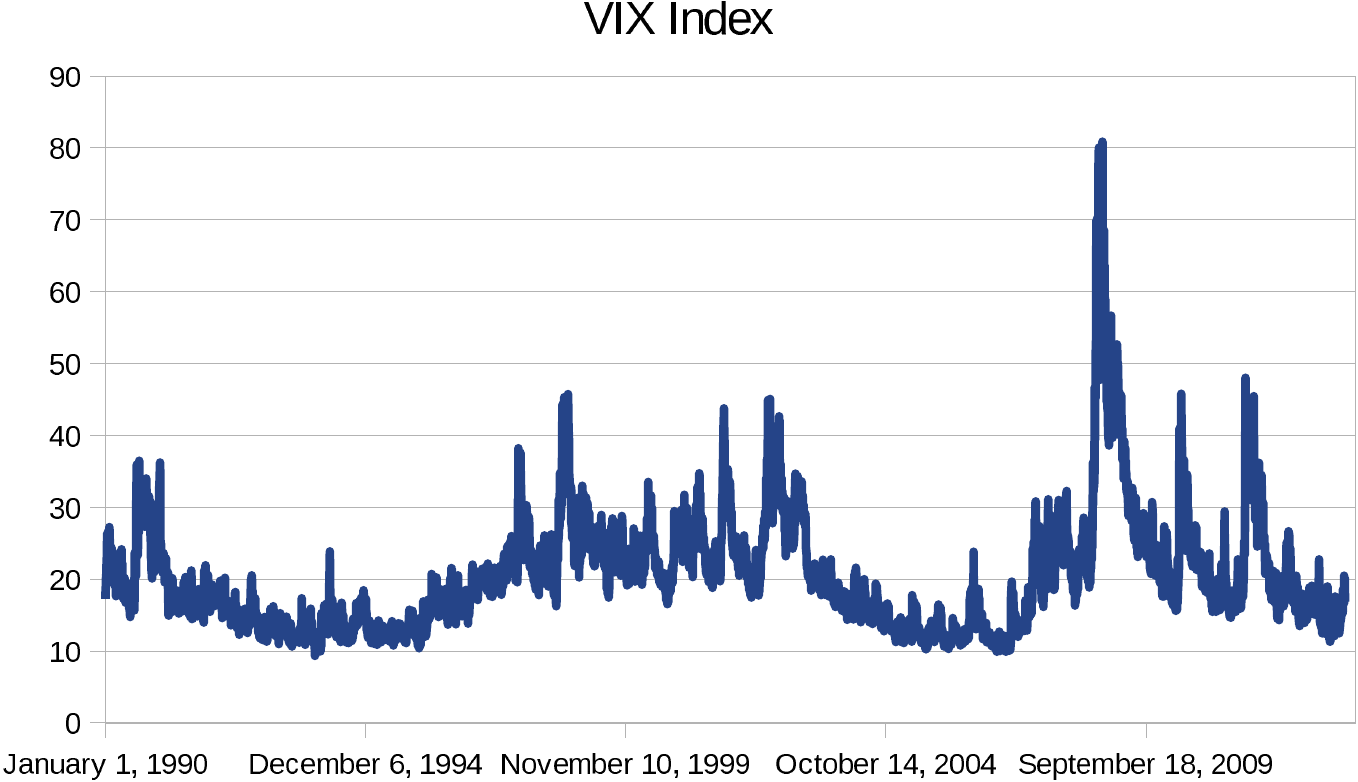}
}
\caption{VIX, CBOE Volatility Index.
Data source: \textit{http://www.cboe.com/micro/VIX/}
}
\label{fig:vix}
\end{figure}

Non-constant volatility is known especially from the market of stocks and the derived options. The most famous index measuring the volatility is arguably VIX, CBOE Volatility Index. It  is a key measure of market expectations of near-term volatility conveyed by S\&P 500 stock index option prices. Since its introduction in 1993, it has been considered by many to be a barometer of investor sentiment and market volatility\footnote{see \textit{http://www.cboe.com/micro/VIX/vixintro.aspx}}. We present its evolution in Figure \ref{fig:vix}.

Moreover, besides the volatility being non-constant and stochastic, there is an evidence that it evolves in a different time scale than the stock price, see a concise book \cite{sircar1} by 
Jean-Pierre Fouque,  George Papanicolaou and K. Ronnie Sircar summarizing their work this area of 
using perturbation methods for the partial differential equation for the option prices in models incorporating this feature. 

Approximately ten years later, in 2011, the same authors and in addition Knut Solna, published a new book \cite{sircar2} with a broader content, 
\textit{Multiscale Stochastic Volatility for Equity, Interest Rate, and Credit Derivatives}, thus featuring 
the topic of interest rates already in the title. The randomness of volatility and interest in its measurement can be seen also from the fact, that CBOE has started to calculate also volatility indices related to interest rates market:
CBOE/CBOT 10-year U.S. Treasury Note Volatility Index\footnote{\textit{www.cboe.com/VXTYN}} and
CBOE Interest Rate Swap Volatility Index\footnote{\textit{www.cboe.com/SRVX}}.

As an example, let us consider stochastic volatility Vasicek model, as given in \cite{sircar1}. It differs from the ordinary Vasicek model by its volatility. Instead of a constant, it is a nonnegative function $f$ evaluated in the value of a stochastic process $y$, following an Ornstein-Uhlenbeck type:
\begin{eqnarray}
d r &=& \kappa_1 (\theta_1 - r) \: d t + f(y) \: d w_1, \nonumber \\
d y &=& \kappa_2 (\theta_2 - y) \: d t + v \:  d w_2, \nonumber 
\end{eqnarray}
where the correlation between the increments $d w_1$ and $d w_2$ is $\rho \in (-1,1)$. Empirical data suggest $\rho>0$, see, e.g.,  \cite[p. 177]{sircar1}.

Another example of a stochastic volatility short rate model has been proposed by Fong and Vasicek in \cite{fong-vasicek} by the following system of stochastic differential equations:
\begin{eqnarray}
d r &=& \kappa_1 (\theta_1 - r) \: d t + \sqrt{y} \:  d w_1, \nonumber \\
d y &=& \kappa_2 (\theta_2 - y) \: d t +  v \sqrt{y} \:  d w_2, \nonumber 
\end{eqnarray}
where again  the Wiener processes can be correlated and the correlation between the increments $d w_1$ and $d w_2$ is $\rho \in (-1,1)$. If the market prices of risk are given by\footnote{Note that this model is a generalization of the one-factor CIR model and the choices for market prices of risk can be seen as generalizations of the said model too.} $\lambda_1 \sqrt{y}$ (market price of risk of short rate) and $\lambda_2 \sqrt{y}$ (market price of risk of volatility), then the partial differential equation for the bond price can be split into solving a system of three ordinary differential equation.

\subsection{Convergence multiple-factor models modeling entry to a monetary union}

\begin{figure}
\centerline{
\includegraphics[width=0.6\textwidth]{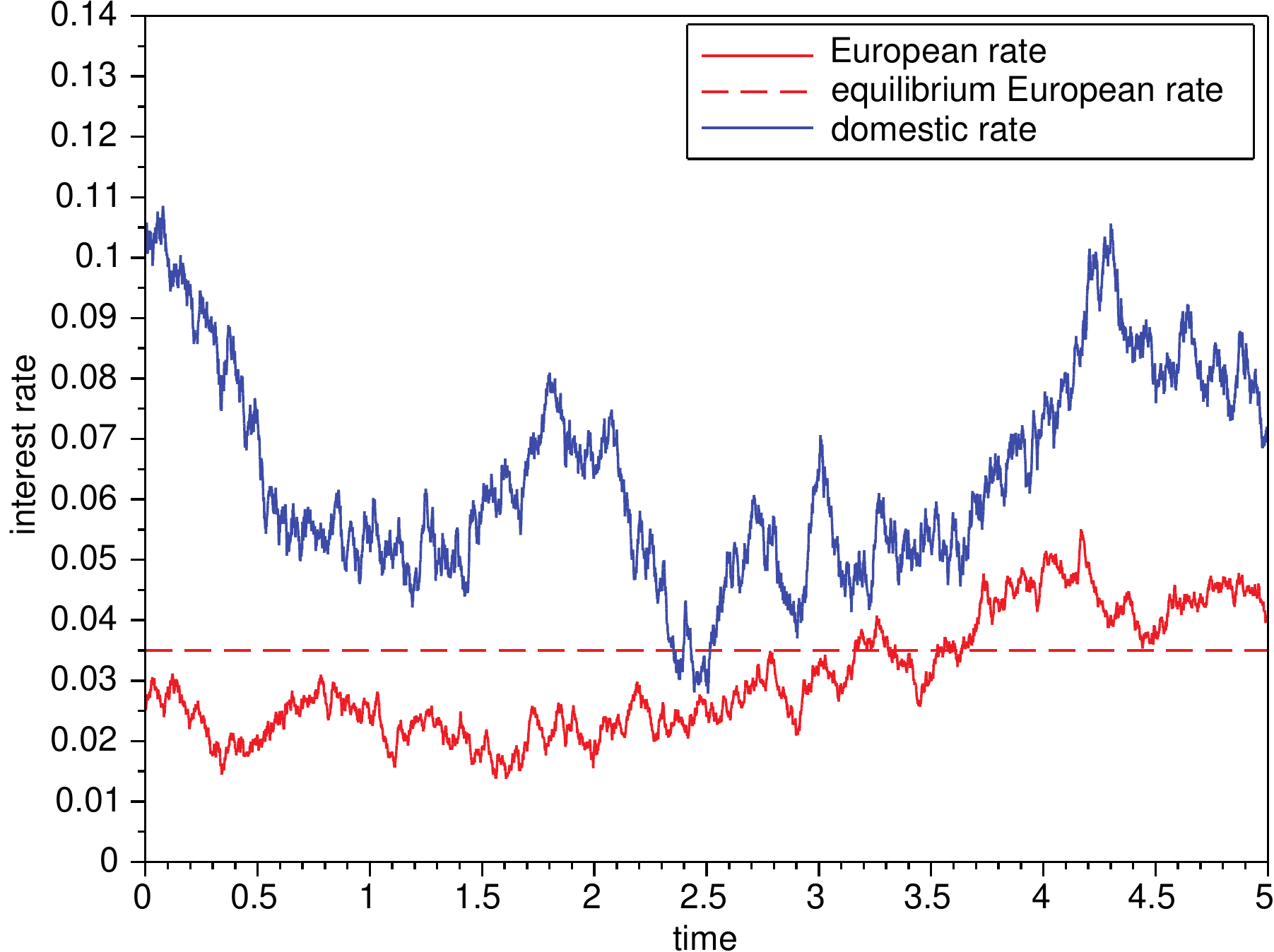}
}
\caption{Sample paths of the European and the domestic short rate in Corzo-Schwarz convergence model with parameters
$a=0.0938, b=3.67, \sigma_d =0.032$, $c=0.2087, d=0.035, \sigma_e=0.016$,  $\rho=0.219$, taken from \cite{corzo-schwarz}. }
\label{fig:conv}
\end{figure}

The basic convergence model of interest rates is suggested by Corzo and Schwarz in \cite{corzo-schwarz}, where they model the interest rates before the formation of the European monetary union. Participating countries fixed their exchange rate to Euro in January 1999. With fixed exchange rate, the interest rates have to be the same across the countries. However, already before fixing the exchange rate, the convergence of the interest rates  in participating countries was possible to be observed. This motivates the following model for the European short rate $r_e$ and the domestic short rate $r_d$:
\begin{eqnarray}
d r_d &=& (a + b(r_e - r_d)) \, d t + \sigma_d \, d w_d, \label{conv-vas-1} \\
d r_e &=& c(d - r_e) \, d t + \sigma_e \, d w_e, \label{conv-vas-2} 
\end{eqnarray}
where the Wiener processes are, in general, correlated: $\textrm{cov}(d w_d, d w_e)=\rho \, d t$. Note that the equation (\ref{conv-vas-2}) is a Vasicek model for the European rate, while (\ref{conv-vas-1}) models a reversion of the domestic rate to the European rate, with a possible minor divergence given by $a$. Figure \ref{fig:conv} shows sample trajectories for the parameters 
$c=0.2087, d=0.035, \sigma_e=0.016$ for the European rate,
$a=0.0938, b=3.67, \sigma_d =0.032$ for the domestic rate and the correlation $\rho=0.219$, taken from \cite{corzo-schwarz}.
Note that in the case of nonzero $a$, the instantaneous drift from (\ref{conv-vas-1}) forces the domestic rate to revert not exactly to the European rate $r_e$, but the value $r_e + a/b$. For the given set of the parameters, the "divergence term" $a/b$ equals to approximately 0.02, which can be observed in Figure \ref{fig:conv}. 
However, with fixed exchange rate, economically plausible value of $a$ is zero. Indeed, when the original model was estimated using the last 3.5 years before entering  the European Monetary Union (EMU) in \cite{corzo-schwarz}, this coefficient turned to be highly insignificant. We also note that negative value of the parameter $a$ would, interestingly, cause also mathematical problems in the generalizations of the model (related to the short rate evolution as well as the bond prices), see \cite{lacko-dp}.

In the market prices of risk are constant, there is an explicit solution for the domestic bond prices\footnote{Note that an explicit solution for the European bonds follows from the fact that we are using a classical Vasicek model for the European interest rates.} of the form
\begin{equation}
P(\tau,r_d,r_e) = A(\tau)e^{-B(\tau) r_d - C(\tau) r_e}. \label{eq:conv-form}
\end{equation}
In \cite{corzo-schwarz} authors claim that the same analysis can be done for the CIR-type convergence model; this has been studied by Lacko in \cite{lacko-dp}. 

If the correlation between $d w_d$ and $d w_e$ is zero, then the solution can be again written in the  form (\ref{eq:conv-form}) and the functions can be found numerically by solving a system of ordinary differential equations. In the general correlated case, the solution cannot be written in the separated form (\ref{eq:conv-form}). This is true also for another natural  generalization, where the European rate is modeled by a CKLS-type process (\ref{eq-ckls-}) and we allow a general form of volatility $\sigma_d r^{\gamma_d}$ also in the equation (\ref{conv-vas-1}) describing the behavior of the domestic rate. An analytical approximation formula for bond prices the CKLS-type model is studied by Z\'ikov\'a and Stehl\'ikov\'a in \cite{zikova-stehlikova}.

\section{Approximate analytical solutions in selected bond pricing  problems}

\fancyhead[EC,OC]{Approximate analytical solutions}
\fancyhead[EL,OR]{\thepage}

Let us consider an example of market interest rates and Euribor rates in particular. Panel banks provide daily quotes of the rate, rounded to two decimal places, that each panel bank believes one prime bank is quoting to another prime bank for interbank
term deposits within the Euro zone. Then, after collecting the data from panel banks: The calculation agent shall, for each maturity, eliminate the highest and lowest 15\% of
all the quotes collected. The remaining rates will be averaged and rounded to three
decimal places. These rates are quoted in percentage points. After dividing them by 100, we obtain them as decimal numbers which are used as the variable $r$ in the models described in the previous chapter. It follows that the value, e.g., 0.123 percentage points from the market data is not an exact figure, but, in terms of decimal numbers, can represent anything from the interval $[0.001225, 0.001235)$. On the other hand, any two numbers  from this interval obtained from models would be in practice indistinguishable. Therefore, going above a certain precision in the computations does not bring any practical advantage when analyzing the market interest rates. In other words, two approximative results that coincide to  certain 
decimal points are practically equally useful and therefore comparing their computational complexity is in place. Approximate analytical solutions, which we deal with, are very convenient in this regard.

\subsection{Chan-Karolyi-Longstaff-Sanders model}

In this section we consider the Chan-Karolyi-Longstaff-Sanders (CKLS hereafter) model in the risk neutral measure
\begin{equation}
 d r=(\alpha+\beta r) \, d t+ \sigma r^{\gamma} \, d w, \label{eq-ckls}
\end{equation}
where $w$ is a Wiener process. 
Note that the linear drift is consistent with the physical measure formulation and choice of market price of risk in the original 
Vasicek model from \cite{vasicek} with $\gamma=0$ and the Cox-Ingersoll-Ross (CIR hereafter) model proposed in \cite{cir} with 
$\gamma=1/2$, see (\ref{eq:pdr:vas}) and (\ref{eq:pdr:cir}).

The price $P(\tau,r)$ of the discount bond, when the current level of the short rate is $r$ and time remaining to maturity is $\tau$, is then given by the solution to the partial differential equation
\begin{equation}
-\partial_{\tau} P + \frac{1}{2} \sigma^2 r^{2 \gamma} \partial^2_{r} P + (\alpha + \beta r)\partial_{r} P - rP =0, \; r>0, \; \tau \in (0,T)
\label{PDE-1f}
\end{equation}
satisfying the initial condition $P(0,r)=1$ for all $r>0$, see, e.g., \cite{kwok}, \cite{brigo-mercurio}. 
Recall that in the case of Vasicek and CIR models the explicit solutions to bond pricing partial differential equations are known.

\subsubsection{Approximation formula due to Choi and Wirjanto}

Consider the stochastic differential equation (\ref{eq-ckls}) in the risk neutral measure for the evolution of the short rate $r$ and the corresponding partial differential equation  (\ref{PDE-1f}) for the bond price $P(\tau,r)$.
The main result of the paper \cite{choi-wirjanto} by Choi and Wirjanto is the following approximation $P^{ap}$ for the exact solution $P^{ex}$:

\begin{theorem}\cite[Theorem 2]{choi-wirjanto}
The approximate analytical solution $P^{ap}$ is given by
\begin{eqnarray}
\label{1f-approximation-formula}
\ln P^{ap}(\tau,r) &=& -rB+\frac{\alpha}{\beta} (\tau-B)+ \left( r^{2 \gamma} +q \tau \right) \frac{\sigma^2}{4 \beta} \left[ B^2 + \frac{2}{\beta} (\tau-B) \right] 
\nonumber \\
& & -q \frac{\sigma^2}{8 \beta^2} \left[B^2(2 \beta \tau-1) - 2B \left(2 \tau - \frac{3}{\beta} \right) + 2 \tau^2 - \frac{6 \tau}{\beta} \right]
\end{eqnarray}
where
\begin{equation}
q(r) = \gamma(2 \gamma -1)\sigma^2 r^{2(2 \gamma-1)} + 2 \gamma r^{2 \gamma-1} (\alpha+\beta r) \label{qr}
\end{equation}
and 
\begin{equation}
B(\tau) = (e^{\beta \tau}-1)/\beta.
\end{equation}
\end{theorem}

The derivation of the formula (\ref{1f-approximation-formula}) is based on calculating the price as an expected value in the risk neutral measure. The tree property of conditional expectation was used and the integral appearing in the exact price was approximated to obtain a closed form approximation. The reader is referred to \cite{choi-wirjanto} for more details of the derivation of  (\ref{1f-approximation-formula}).

Authors furthermore showed that such an approximation coincides with the exact solution in the case of the Vasicek model \cite{vasicek}. Moreover, they compared the above approximation with the exact solution of the CIR model which is also known in a closed form. 
Graphical demonstration of relative mispricing, i.e., the relative error in the bond prices, has been also provided by the authors. 

\subsubsection{Asymptotic analysis of the Choi and Wirjanto approximation formula }

As it can be seen in the numerical examples given in \cite{choi-wirjanto}, the error in bond prices is smaller in the case of $\tau$ small. Also, for $\tau=0$ the formula is exact. This suggests using $\tau$ as a small parameter in the asymptotic analysis. 

Using the exact solution $P^{ex}_{CIR}$ in the case of $\gamma=1/2$ (i.e., the Cox-Ingersoll-Ross model), computing its expansion in $\tau$ around the point $\tau=0$ and comparing it with the expansion of the Choi and Wirjanto approximate formula $P^{ap}_{CIR}$ with  $\gamma=1/2$  we obtain
$$
\ln P^{ap}_{CIR}(\tau,r) - \ln P^{ex}_{CIR}(\tau,r) = -\frac{1}{120} \sigma^2 \left[ \alpha \beta + r(\beta^2 - 4 \sigma^2) \right] \tau^5 + o(\tau^5)
$$
as $\tau \rightarrow 0^+$. Considering logarithms of the bond prices enables us to estimate the relative error in the bond prices 
(the relative mispricing from the previous subsection) and the absolute error in the interest rates forming a term structure of interest rate.

The result of expanding the approximate and exact solutions in the case of the CIR model motivates finding a similar estimate also in the case of a general CKLS model, i.e., for arbitrary $\gamma$. In the paper \cite{stehlikova-sevcovic} we proved the following theorem:

\begin{theorem} \cite[Theorem 3]{stehlikova-sevcovic}
\label{theorem-accuracy-for-ln-wirjanto}
Let $P^{ap}$ be the approximative solution given by (\ref{1f-approximation-formula}) and $P^{ex}$ be the exact bond price given as a unique complete solution to (\ref{PDE-1f}). Then
\[
\ln P^{ap}(\tau,r) - \ln P^{ex}(\tau,r) = c_5(r) \tau^5 + o(\tau^5) 
\]
as $\tau \rightarrow 0^+$ where
\begin{eqnarray}
\label{c5(r)wirjanto}
c_5(r) &=& -\frac{1}{120} \gamma r^{2(\gamma-2)} \sigma^2 
\left[ 
2 \alpha^2 (-1+2 \gamma) r^2 + 4 \beta^2 \gamma r^4 
- 8 r^{3+2\gamma} \sigma^2
\right.
\nonumber \\ 
& &
+ 2 \beta (1-5 \gamma + 6 \gamma^2) r^{2 (1+\gamma)} \sigma^2 
+\sigma^4 r^{4 \gamma} (2\gamma-1)^2 (4 \gamma-3) 
 \\
& &
\left.
+ 2 \alpha r \left( \beta(-1+4 \gamma)r^2 + (2 \gamma -1)(3\gamma -2) r^{2 \gamma} \sigma^2 \right)
 \right]. \nonumber
\end{eqnarray}
\end{theorem}

Moreover, the method of the proof enabled to propose an approximation formula of a higher accuracy, as stated in the following theorem.

\begin{theorem} \cite[Theorem 4]{stehlikova-sevcovic}
\label{theorem-accuracy2-for-ln-wirjanto}
Let $P^{ex}$ be the exact bond price. Let us define an improved approximation $P^{ap2}$ by the formula 
\begin{equation}
\ln P^{ap2}(\tau,r)  = \ln P^{ap}(\tau,r) - c_5(r) \tau^5 - c_6(r) \tau^6
\label{approx-higher-order}
\end{equation}
where $\ln P^{ap}$ is given by (\ref{1f-approximation-formula}), $c_5(\tau)$ is given by (\ref{c5(r)wirjanto}) in Theorem 
\ref{theorem-accuracy-for-ln-wirjanto}  and
\[
c_6(r)=\frac{1}{6} \left( \frac{1}{2} \sigma^2 r^{2 \gamma} c_5''(r) + (\alpha+\beta r) c_5'(r) - k_5(r) \right)
\]
where $c_5^\prime$ and $c_5^{\prime\prime}$ stand for the first and second derivative of $c_5(r)$ w. r. to $r$ and $k_5$ is defined by
\begin{eqnarray}
\label{k5}
k_5(r)&=&
\frac{\gamma\sigma^2}{120}
 r^{2 \left( -2 + {\gamma} \right) }  
    \left[ 6 {\alpha}^2 \beta \left( -1 + 2 {\gamma} \right)  r^2 + 12 {\beta}^3 {\gamma} r^4 
 - 10 {\left( 1 - 2 {\gamma} \right) }^2 r^{1 + 4 {\gamma}} {\sigma}^4
\right. \nonumber \\
&&
+ 6 {\beta}^2  \sigma^2\left( 1 - 5 {\gamma} + 6 \gamma^2 \right)  r^{2 \left( 1 + {\gamma} \right) } 
 \nonumber \\
&&
 + \beta r^{2 \gamma}\sigma^2 \left( -10 \left( 5 + 2 \gamma \right)  r^3 +  
         3 {\left( 1 - 2 {\gamma} \right) }^2 \left( -3 + 4 {\gamma} \right)  r^{2 {\gamma}} {\sigma}^2 \right)  \nonumber \\
&& 
+ 2 \alpha r \biggl( 3 {\beta}^2 \left( -1 + 4 {\gamma} \right)  r^2 + 
         3 \beta \left( 2 - 7 {\gamma} + 6 {{\gamma}}^2 \right)  r^{2 {\gamma}} {\sigma}^2 
\nonumber \\
&& 
\qquad - \left. 5 \left( -1 + 2 {\gamma} \right)  r^{1 + 2 {\gamma}} {\sigma}^2 \biggr)  \right]\,.
\end{eqnarray}
Then the difference between the higher order approximation $\ln P^{ap2}$ given by (\ref{approx-higher-order}) and the exact solution $\ln P^{ex}$ satisfies
$$
\ln P^{ap2}(\tau,r) - \ln P^{ex}(\tau,r) = o(\tau^6)
$$
as $\tau \rightarrow 0^+$.
\end{theorem}

In Table \ref{irmodels-tab2}  we show $L_{\infty}$ and $L_2\,-\,$norms\footnote{$L_{p}$ and $L_{\infty}$ norms of a function $f$ defined on a grid  with step $h$ are given by $\Vert f \Vert_p=\left( h \sum |f(x_i)|^p \right)^{1/p}$ and $\Vert f \Vert_{\infty}=\max |f(x_i)|$.} with respect to $r$ of the difference $\ln P^{ap} - \ln P^{ex}$ and $\ln P^{ap2} - \ln P^{ex}$ where we considered $r \in [0,0.15]$. We also compute the experimental order of convergence (EOC) in these norms. Recall that the experimental order of convergence gives an approximation of the exponent $\alpha$ of expected power law estimate for the error $\Vert\ln P^{ap}(\tau,.) - \ln P^{ex}(\tau,.)\Vert = O(\tau^{\alpha})$ as $\tau \rightarrow 0^+$. The $EOC_i$ is given by a ratio
\[
EOC_i = \frac{\ln (err_i/err_{i+1})}{\ln ( \tau_i/\tau_{i+1})},
\quad\hbox{where }\ \ err_i = \Vert\ln P^{ap}(\tau_i,.) - \ln P^{ex}(\tau_i,.)\Vert_p\,.
\]

\begin{table}
\caption{The $L_{\infty}$ and $L_2\,-\,$errors for the original $\ln P_{CIR}^{ap}$ and improved $\ln P_{CIR}^{ap2}$ approximations. Parameters are set to be equal to $\alpha=0.00315$, $\beta=-0.0555$, $\sigma=0.0894$.  Source: Stehl\'{\i}kov\'{a} and \v{S}ev\v{c}ovi\v{c} \cite{stehlikova-sevcovic}.}
\begin{center}

\small
\begin{tabular}{c|cc|cc} 
$\tau$ &  $\Vert\ln P^{ap} - \ln P^{ex}\Vert_{\infty}$ &  EOC &  $\Vert\ln P^{ap2} - \ln P^{ex}\Vert_{\infty}$ & EOC\\
\hline
1 &  $ 2.774 \times 10^{-7}$ &  4.930 &$4.682 \times 10^{-10}$  & 7.039 \\ 
0.75 &  $6.717 \times 10^{-8}$ &  4.951& $ 6.181 \times 10^{-11}$& 7.029 \\ 
0.5 &  $9.023 \times 10^{-9}$ &  4.972 &$3.576 \times 10^{-12}$& 7.004 \\ 
0.25 & $2.876 \times 10^{-10}$  &  --  &$2.786 \times 10^{-14}$& -- \\ 
\hline
\end{tabular}
\end{center}

\begin{center}
\vglue 2mm
\small
\begin{tabular}{c|cc|cc} 
$\tau$ &  $\Vert\ln P^{ap} - \ln P^{ex}\Vert_{2}$ &  EOC &  $\Vert\ln P^{ap2} - \ln P^{ex}\Vert_{2}$ & EOC\\
\hline
1 &  $6.345 \times 10^{-8}$ &  4.933 & $9.828 \times 10^{-11}$ & 7.042\\ 
0.75 & 1.535 $\times 10^{-8}$ &  4.953 & $1.296 \times 10^{-11}$ & 7.031\\ 
0.5 & 2.061 $\times 10^{-9}$ & 4.973 & $7.492 \times 10^{-13}$ & 7.012\\ 
0.25 &  6.563 $\times 10^{-11}$ &  --  & $5.805 \times 10^{-15}$ & -- \\
\hline
\end{tabular}
\end{center}

\label{irmodels-tab2}
\end{table} 

In Table~\ref{irmodels-tab2}  we show the $L_2\,-\,$error of the difference between the original and improved approximations for larger values of $\tau$. It turns out that the higher order approximation $P^{ap2}$ gives about twice better approximation of bond prices in the long time horizon up to 10 years.

\subsubsection{Approximation based on the Vasicek model}

Our aim is to propose a formula which is as simple as possible, but still yields a good approximation to the exact bond prices. Using an approximation in calibration of the model requires many evaluations of its value for different sets of parameters, as well as times to maturity and the short rate levels. Therefore, its simple form can increase the efficiency of the calibration procedure. In particular, the approximation published by Stehl\'ikov\'a in \cite{stehlikova} presented in this section leads to a one-dimensional optimization problem.

Again, we consider the model 
(\ref{eq-ckls}) in the risk neutral measure for the evolution of the short rate $r$ and the corresponding partial differential equation  (\ref{PDE-1f}) for the bond price $P(\tau,r)$.

Recall that
in the case of Vasicek model, i.e., for $\gamma=0$, the solution $P_{vas}$ can be expressed in the closed form:
\begin{equation}
\ln P_{vas}(\tau,r) = \left( \frac{\alpha}{\beta} + \frac{\sigma^2}{2 \beta^2} \right)
\left(\frac{1-e^{\beta \tau}}{\beta}+\tau \right)
 + \frac{\sigma^2}{4 \beta^3} (1-e^{\beta \tau})^2 + \frac{1-e^{\beta \tau}}{\beta} r.
\label{vasicek-price}
\end{equation}
Now, let us consider a general model (\ref{eq-ckls}) and the approximation of the bond price obtained by substituting the instantaneous volatility $\sigma r^{\gamma}$ for $\sigma$ in the Vasicek price (\ref{vasicek-price}), i.e.,
\begin{equation}
\ln P^{ap}(\tau,r) = \left( \frac{\alpha}{\beta} + \frac{\sigma^2 r^{2 \gamma}}{2 \beta^2} \right)
\left(\frac{1-e^{\beta \tau}}{\beta}+\tau \right)
 + \frac{\sigma^2 r^{2 \gamma}}{4 \beta^3} (1-e^{\beta \tau})^2 + \frac{1-e^{\beta \tau}}{\beta} r.
\label{ckls-app-price}
\end{equation}

\begin{theorem}\cite[Theorem 1]{stehlikova}
\label{theorem-accuracy-for-ln}
Let $P^{ap}$ be the approximate solution given by (\ref{ckls-app-price}) and $P^{ex}$ be the exact bond price given as a solution to (\ref{PDE-1f}). Then
\[
\ln P^{ap}(\tau,r) - \ln P^{ex}(\tau,r) = c_4(r) \tau^4 + o(\tau^4) 
\]
as $\tau \rightarrow 0^+$ where
\begin{eqnarray}
\label{c5(r)}
c_4(r) &=& -\frac{1}{24} \gamma r^{2 \gamma -2} \sigma^2 [2 \alpha r + 2 \beta r^2 + (2 \gamma -1) r^{2 \gamma} \sigma^2].
\nonumber 
\end{eqnarray}
\end{theorem}

For the practical usage of the approximate formula, besides the order of accuracy  also the absolute value of the error is significant. 

\begin{figure}
  \centerline{
    \includegraphics[width=0.8 \textwidth]{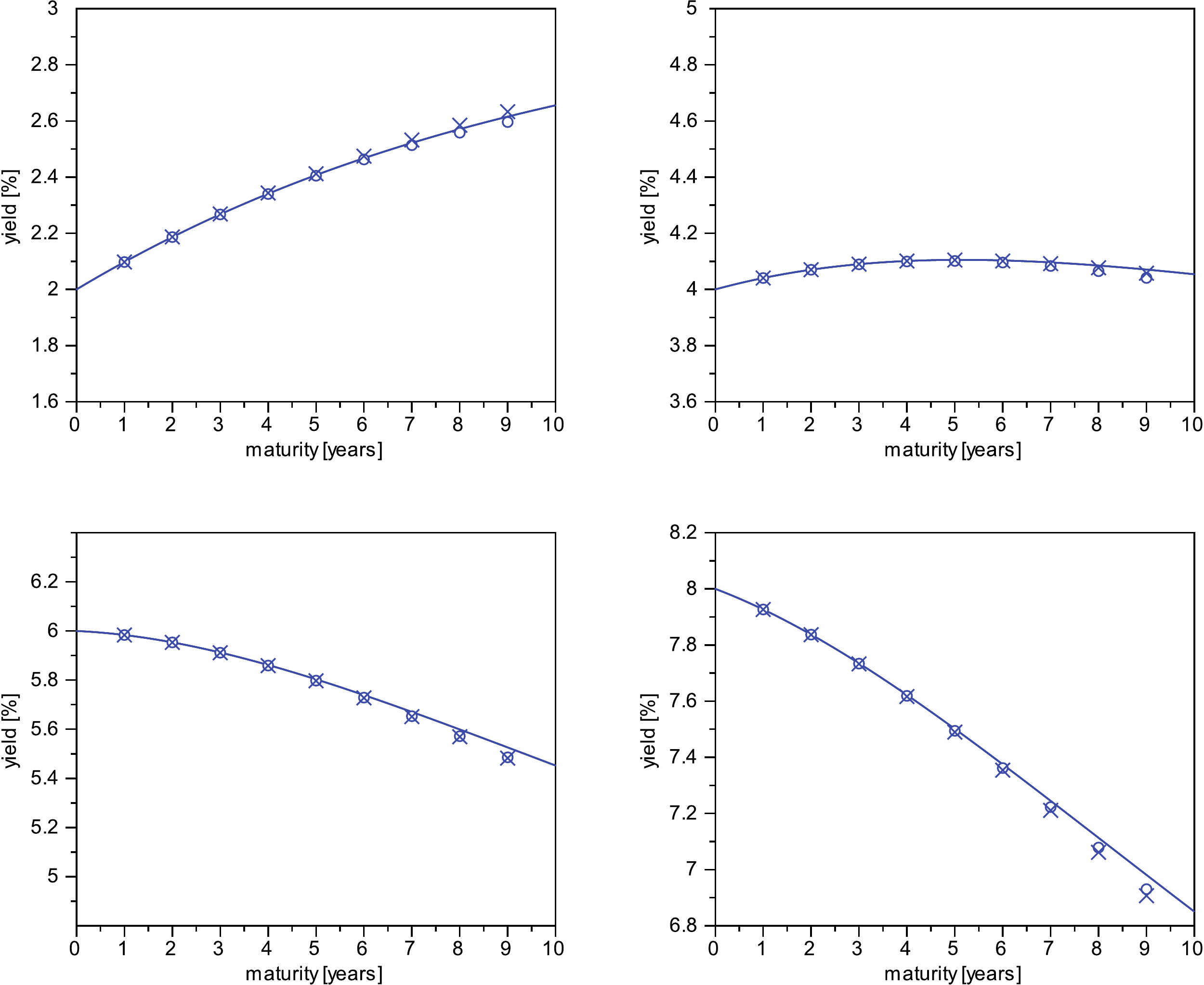}
  }
\caption{Comparison of the exact term structures in the CIR model (solid line), approximation based on Vasicek model from \cite{stehlikova} by Stehl\'ikov\'a (crosses) and approximation from \cite{choi-wirjanto} by Choi and Wirjanto (circles). Parameters are set to be equal to $\alpha=0.00315$, $\beta=-0.0555$, $\sigma=0.0894$. Source: Stehl\'ikov\'a, \cite{stehlikova}.}
\label{fig:approx-term-structures}
\end{figure}

Comparison of the approximation with the exact values in the case of CIR models and parameter values from \cite{choi-wirjanto}
show (cf. \cite{stehlikova} for the exact figures) that for shorter  maturities  the differences are less than the accuracy to which the market data are quoted. Euribor, for example, is quoted in percentage points rounded to three decimal places. Moreover, Figure \ref{fig:approx-term-structures} shows that even though the accuracy of this approximation is one order lower to that of the approximation from \cite{choi-wirjanto}, it gives numerically comparable results for the real set of parameters.
 
Let us consider the calibration of the one-factor model based on the comparison of theoretical and market interest rates, where the parameters are chosen to minimize the function
\begin{equation}
F=\frac{1}{mn} \sum_{i=1}^n \sum_{j=1}^m w_{ij} \left( R(\tau_j,r_i) - R_{ij} \right)^2, \label{ucelova-funkcia}
\end{equation}
where  $r_i$ ($i=1,\dots,n$) is the short rate observed on the $i$-th day, $\tau_j$ ($j=1,\dots,m$) is the $j$-th maturity of the interest rates in the data set, $R_{ij}$ is the interest rate with maturity $\tau_j$ observed on i-th day, $R(\tau,r)$ is the interest rate with maturity $\tau$ corresponding to the short rate $r$ computed from the model with the given parameters and $w_{ij}$ are the weights. In \cite{sevcovic-csajkova-1} and \cite{sevcovic-csajkova-2}, this approach was used with $w_{ij}=\tau_j^2$ (i.e., giving more weight to fitting longer maturities) to calibrate Vasicek and CIR models using the explicit solutions for interest rates. To achieve the global minimum of the objective function, the authors used evolution strategies. 

If we attempted to use this method to estimate a  model with different $\gamma$ without analytical approximation, it would become computationally demanding, since each evaluation of the objective function  would require   numerical solutions of the PDE (\ref{PDE-1f}). Note that the evaluation is needed for every member of the population in the evolution strategy (see \cite{sevcovic-csajkova-2} for   details).  Using  an analytical approximation simplifies the computation of the objective function, but in general the dimension of the optimization problem is unchanged. We show that using the approximation proposed in this paper, we are able to reduce the calibration  to a one-dimensional optimization problem   which can be quickly solved using simple algorithms.

Hence we consider the criterion (\ref{ucelova-funkcia}) with replacing $R(\tau,r)$ by its approximation $R^{ap}(\tau,r)$ calculated from (\ref{ckls-app-price}). Note that the approximation formula $\ln P^{ap}$ is a linear function of parameters $\alpha$ and $\sigma^2$; it can be written as
\[
\ln P^{ap} (\tau,r)=  c_0(\tau,r) + c_1(\tau,r) \alpha + c_2(\tau,r) \sigma^2,
\]
where
\[ c_0 = \frac{1 - e^{\beta \tau}}{\beta} r, \; c_1 = \frac{1}{\beta} \left( \frac{1 - e^{\beta \tau}}{\beta} + \tau \right), \;
 c_2 = \frac{r^{2 \gamma}}{2 \beta^2} \left( \frac{1- e^{\beta \tau}}{\beta} + \tau + \frac{ \left(1 - e^{\beta \tau} \right)^2}{2 \beta}
 \right).
\]
Hence taking the derivatives of (\ref{ucelova-funkcia}) with respect to $\alpha$ and $\sigma^2$ and setting them equal to zero leads to a system of linear equations for these two parameters.
It means that once we fix $\gamma$ and treat $\beta$ as parameter, we obtain the corresponding optimal values of $\alpha$ and $\sigma^2$ for each $\beta$. Substituting them into (\ref{ucelova-funkcia}) then leads to a one-dimensional optimization problem. Doing this over a  range of values of $\gamma$ allows us to find the optimal parameter $\gamma$ as well.

We show the proposed idea on simulated data. Once again, we consider the CIR model with parameters from \cite{choi-wirjanto} and   
simulate the daily 
 term structures - interest rates with maturities of $1, 2, 3, \dots, 12$ months  using the exact formula for CIR model - for a period of year. In the objective function (\ref{ucelova-funkcia}) we use the weights $w_{ij}=\tau_j^2$ as in  \cite{sevcovic-csajkova-1} and  \cite{sevcovic-csajkova-2}. Afterwards we repeat the same procedure with maturities of 1, 2, 3, 4 and 5 years.

Results of the estimation   for several values of $\gamma$ are presented in Table \ref{tab-calibration}, we show the estimated parameters and the optimal value of the objective function $F$.  

\begin{table}
\caption{Estimation of the parameter $\gamma$ using the approximate formula for interest rates. The data were simulated using the exact formula with the parameters $\alpha=0.00315$, $\beta=-0.0555$, $\sigma=0.0894$, $\gamma=0.5$. Maturities used were $1,2,\dots,12$ months (above) and $1,2,\dots,5$ years (below). Source: Stehl\'ikov\'a, \cite{stehlikova}}
\begin{center}
\vglue 2mm
\small
\begin{tabular}{c|ccc|c} 
$\gamma$ & $\alpha$ & $\beta$ &  $\sigma$ &  optimal value of $F$ \\
\hline 
0    & 0.00324 & -0.0578  &  0.0176 &  1.1 $\times 10^{-12}$  \\
0.25 & 0.00319 & -0.0565  &  0.0403 &  2.9 $\times 10^{-13}$  \\
0.5  & 0.00315 & -0.0555  &  0.0896 &  1.1 $\times 10^{-15}$  \\
0.75 & 0.00312 & -0.0548  &  0.1912 &  6.3 $\times 10^{-13}$  \\
1    & 0.00310 & -0.0548  &  0.3813 &  2.5 $\times 10^{-12}$  \\
\hline
\end{tabular}
\end{center}

\begin{center}
\vglue 2mm
\small
\begin{tabular}{c|ccc|c} 
$\gamma$ & $\alpha$ & $\beta$ &  $\sigma$ &  optimal value of $F$ \\
\hline 
0    & 0.00377 &  -0.0663 & 0.0214 & 1.0 $\times 10^{-8}$  \\
0.25 & 0.00344 &  -0.0607 & 0.0432 & 2.4 $\times 10^{-9}$  \\
0.5  & 0.00311 &  -0.0553 & 0.0860 & 2.2 $\times 10^{-10}$  \\
0.75 & 0.00281 &  -0.0506 & 0.1688 & 6.7 $\times 10^{-9}$  \\
1    & 0.00256 &  -0.0471 & 0.3238 & 2.7 $\times 10^{-8}$  \\
\hline
\end{tabular}
\end{center}

\label{tab-calibration}
\end{table}

\subsection{General one-factor models: power series expansions}
The approximations considered in the previous sections share a common feature:
their order of accuracy can be expressed in the form
\begin{equation}
\ln P^{ap}(\tau,r) - \ln P (\tau,r) = c(r) \tau^{\omega} + o(\tau^{\omega}) \label{eq-order}
\end{equation}
as $\tau \rightarrow 0^+$, where $P$ is the exact bond price and $P^{ap}$ is the proposed approximation. The relation (\ref{eq-order}) asserts that the Taylor series of $\ln P^{ap}$ and $\ln P$ coincide up to the certain order. In particular, in \cite{stehlikova-sevcovic} it has been shown that for the formula for CKLS model from  \cite{choi-wirjanto} the relation (\ref{eq-order}) holds with $\omega=5$  and an improvement leading to $\omega=7$ has been derived. In \cite{stehlikova} a simple formula with $\omega=4$ has been proposed. Similar estimates hold in the case of multi-factor models. These results suggest that the Taylor expansion (either of the price itself and its logarithm) could be a good approximation too.

Let us consider a
general one-factor model with constant coefficients
\begin{equation} 
d r = \mu(r) \, d t + \sigma(r)  \, d w \label{eq:sde2}.
\end{equation}
Recall that the price of the bond $P(\tau,r)$ is a solution to the partial differential equation
\begin{equation}
-\partial_{\tau} P + \mu(r) \partial_{r} P + \frac{1}{2} \sigma^2(r) \partial^2_{r} P - rP=0  \label{eq:pde-tau}
\end{equation}
for all $r>0$, $\tau \in (0,T)$ and the initial condition $P(0,r)=1$ for all $r>0$. Easy transformation of the PDE leads to the equation which is satisfied by the logarithm of the bond price, i.e., $f(\tau,r)= \log P(\tau,r)$:
\begin{equation}
-\partial_{\tau} f = \frac{1}{2} \sigma^2(r) \left[ (\partial_{r} f)^2 + \partial^2_{rr} f \right] + \mu(r) \partial_{r} f - r =0
\label{eq-f}
\end{equation}
for all $r>0$, $\tau \in (0,T)$ and the initial condition $f(0,r)=0$ for all $r>0$.  Writing these functions in series expansions around $\tau=0$ in the form
\begin{equation}
P(\tau,r) = \sum_{j=0}^{\infty} c_j(r) \tau^j, f(\tau,r) = \sum_{j=0}^{\infty} k_j(r) \tau^j \label{eq:p_f_series}
\end{equation}
enables us to compute the parameters $c_j$ or $k_j$ recursively in the closed form.
A practical usage of this approach is determined by the speed of convergence of these series for reasonable values of $\tau$ and $r$. 
Then, we can approximate the bond prices and their logarithms by terminating the infinite sums (\ref{eq:p_f_series}) at a certain index $J$.

We show the results from \cite{stehlikova:taylor}. Firstly, the approximation is tested on CKLS model with the same parameters as in the previous chapter; the results suggest the possibility of practical usage of the proposed matter. 
As an another example, the Dothan model is considered.
The Dothan model \cite{irmodels-dothan} assumes that the short rate in the risk neutral measure follows the stochastic differential equation
$$dr = \mu r dt + \sigma r dw.$$
The zero-coupon bond in the Dothan model has an  explicit solution, but it is computationally complicated (cf. \cite{brigo-mercurio}).
Therefore, we use the Dothan bond prices computed in \cite{hansen-jorgensen} for which the error estimate is available. They are accurate to the given four decimal digits.

Setting $\mu(r)=\mu r$ and $\sigma(r)=\sigma r$ into the recursive formulae for coefficients results in the coefficients for the price and its logarithm. In the numerical experiments we use the values from \cite{hansen-jorgensen}. The authors price zero coupon bonds which pays 100 USD at maturity $T$ (hence its price is 100 times the value considered so far).
Using their iterative algorithm, for $\tau=1,2,3,4,5,10$ they obtain the accuracy to four decimal digits for all combinations of parameters and in several cases  also  for higher maturities.
Selected values from \cite{hansen-jorgensen} are used to test the approximation for a wider range of parameters and maturities, as shown in Table  \ref{tab-dothan-2}.

\begin{table}
\caption{{Bond prices in the Dothan model with indicated parameters and maturities, and the initial value of the short rate 
 $r_0=0.035$ - comparison of Taylor approximation with exact values. Source Stehl\'{\i}kov\'a \cite{stehlikova:taylor}}}

\centering
\small
\smallskip
\begin{tabular}{c|c|ccccc}
parameters & $\tau$ &  Taylor, J=3 &  Taylor, J=5 &  Taylor, J=7 & exact \cite{hansen-jorgensen}  \\ 
\hline
$\mu=0.005$, $\sigma^2=0.01$& 1 & 96.5523 &96.5523   & 96.5523 & 96.5523 \\ 
 & 2 & 93.2082 & 93.2082   & 93.2082 & 93.2082 \\ 
& 3 & 89.9666 & 89.9663   & 89.9663 & 89.9663 \\ 
& 4 & 86.8260 & 86.8251   & 86.8251 & 86.8251 \\ 
& 5  & 83.7852 & 83.7830   & 83.7830 & 83.7830 \\ 
&  10 & 70.0312  & 69.9977   & 69.9982 & 69.9982 \\ 
\hline 
$\mu=0.005$, $\sigma^2=0.02$ & 1 & 96.5525 & 96.5525   & 96.5525 & 96.5525 \\ 
 & 2 &  93.2099 & 93.2098   & 93.2098 &  93.2098 \\  
& 3  & 89.9721 & 89.9715   & 89.9715 &  89.9715 \\  
& 4  & 86.8391 & 86.8370   & 86.8370 & 86.8370  \\
& 5  & 83.8362 & 83.8056   & 83.8057 & 83.8057\\
&  10  & 70.4396 & 70.1530   & 70.1551 &  70.1551 \\
\hline
$\mu=0.005$, $\sigma^2=0.03$ & 1 & 96.5527 & 96.5527   & 96.5527 & 96.5527 \\ 
 & 2 &  93.2115 & 93.2113   & 93.2113 &  93.2113 \\  
& 3  & 89.9776 & 89.9767   & 89.9767 &  89.9767 \\  
& 4  & 86.8521 & 86.8491  & 86.8491 & 86.8491  \\
& 5  & 83.8362 & 83.8287   & 83.8287 & 83.8287 \\
&  10  & 70.4396 & 70.3112   & 70.3151 &  70.3151 \\
\hline
\end{tabular} 

\label{tab-dothan-2} 
\end{table}

The idea of the short time asymptotic expansion can be enhanced, by considering the so-called {exponent expansion}  to derive a closed-form short-time approximation of the Arrow-Debrew prices, from which the prices of bonds or other derivatives can be obtained by a single integration.  This technique, originally introduced in chemical physics by Makri and Miller \cite{makri}, was 
introduced to finance by Capriotti \cite{ee}. In  \cite{stehlikova-capriotti} by Stehl\'ikov\'a and Capriotti, it was employed to compute the bond prices in the Black-Karasinski model.

The exponent expansion is derived for the bond prices in short rate models with $r=r(x)$, where the auxiliary process has the form 
\begin{equation}
d x(t) = \mu(x) \,d t + \sigma \,d w,
\end{equation}
where $\mu(x)$ is a drift function. Note that the process has a 
constant volatility $\sigma$, but in the general case it is possible to map to this case a general state 
dependent volatility function by means of  an integral transformation. Note that this transformation is used also by A\"it-Sahalia in
\cite{ait-sahalia-transition} in his approximation of transition densities.

The bond prices are not computed directly; instead, so-called Arrow-Debreu prices   are approximated by a closed form formula and the bond prices are obtained by a single numerical integration. 
The Arrow-Debreu prices $\psi(x,T;x_0)$ are for each $x_0$ given as solutions to the 
partial differential equation (see  \cite{shreve})
\begin{equation}\label{fp}
\partial_{t} \psi = \Big( -r(x) - \partial_x \mu(x) + \frac{1}{2}\sigma^2 \partial_x^2 \Big) \psi,  
\end{equation}
with the initial condition $\psi(x, 0; x_0) = \delta(x-x_0)$.
Looking for the solution in the form
\begin{equation}
\psi(x, t; x_0) = \frac{1}{\sqrt{2\pi\sigma^2 t}} 
\exp {\left[ -\frac{(x-x_0)^2}{2\sigma^2 t}-W(x,t;x_0)\right]},
\label{ansatz}
\end{equation}
and inserting it into (\ref{fp}) leads to a partial differential equation for $W(x, t; x_0)$. Writing it in the form 
\begin{equation}
W(x, t; x_0) = \sum_{n=0}^{\infty} W_n(x;x_0)\, t^n~,
\label{w}
\end{equation}
allows a recursive computation of the functions $W_n(x;x_0)$ as solutions to 
first order linear ordinary differential equations.

This form of expansion for the bond prices results in a more rapid convergence especially for longer maturities, compared with the simple Taylor expansion described previously, see Table \ref{tab:ee-taylor}.

\begin{table}

\caption{{ Comparison of successive approximations of the bond price with six-months (left) and one-year (right) maturity in Black-Karasinski model with parameters $a=1$, $b=\ln 0.04$, $\sigma=0.85$, when the initial level of the short rate is $r=0.06$. }}

\vskip 0.2cm 

\centering \small
\begin{tabular}{c|cc|cc}

%
order & Taylor & exponent expansion   & Taylor & exponent expansion  \\
\hline
1 & 0.970000 & 0.969249  & 0.940000 & 0.937431 \\
2 & 0.968045 & 0.968138  & 0.932179 & 0.933037 \\
3 & 0.968123 & 0.968140  & 0.932807 & 0.933077 \\
4 & 0.968141 & 0.968142  & 0.933097 & 0.933105 \\
5 & 0.968142 & 0.968142  & 0.933118 & 0.933106 \\
6 & 0.968142 & 0.968142  & 0.933110 & 0.933106 \\
\hline
\end{tabular} 

\label{tab:ee-taylor} 
\end{table}

An  important advantage that separates the exponential expansion
is the possibility to systematically improve its accuracy over large time horizons by means of the convolution approach, cf. \cite{stehlikova-capriotti} for the algorithm. 
  This allows to produce  results accurate to more than 4 significant digits 
even for zero  coupon bonds with maturities over 20 years. This is documented in Table \ref{tab:stehlikova-capriotti} where the results are compared with Monte Carlo prices.

\begin{table}

\caption{{Bond prices computed with the 6th order Exponent Expansion and different convolution steps in Black-Karasinski model with parameters $a=1$, $b=\ln 0.04$, $\sigma=0.85$, when the initial level of the short rate is $r=0.06$, compared with the price obtained by Monte Carlo. Source: Stehl\'ikov\'a and Capriotti, \cite{stehlikova-capriotti}.}}

\vskip 0.2cm 
\centering
\small 

\begin{tabular}{c|ccc|c}

maturity & convolution step: 5 &  convolution  step: 2.5    &  convolution  step: 1 & MC  \\
\hline
5 & 0.65949 & 0.65955  & 0.65966 & 0.6597 \\
10 & 0.46139 & 0.46222  & 0.46229 & 0.4623 \\
20 & 0.26812 & 0.26827  & 0.26831 & 0.2683 \\
\hline
\end{tabular}

\label{tab:stehlikova-capriotti} 
\end{table}

\subsection{Fast time scale of volatility in stochastic volatility models}

In the  paper \cite{stehlikova-sevcovic-kybernetika} by Stehl\'ikov\'a and \v Sev\v covi\v c,  studied a generalized CIR model with a stochastic volatility. The instantaneous interest rate (short rate) $r$ is modeled by the mean reverting process of the form (\ref{eq:cir_sde}) where the constant $\sigma$ appearing in the volatility function $\sigma \sqrt{r}$  is replaced by a square root of a stochastic dispersion $y$, i.e. 
\begin{equation}
d r = \kappa(\theta - r) \, d t + \sqrt{y} \sqrt{r} \, d w_r\,.
\label{CIR-short-rate-stoch}
\end{equation}
The stochastic differential equation for the short rate is given by 
\begin{equation}
d y=\alpha(y) \, d t + v \sqrt{y} \, d w_y,
\label{slowly-osc}
\end{equation}
with certain conditions given on the function $\alpha\!\!: [0,\infty) \rightarrow \mathbb{R}$ at zero and infinity, see \cite[Assumption A]{stehlikova-sevcovic-kybernetika} and a concrete example\footnote{The concrete example of  a function $\alpha$ considered in the paper models a volatility clustering phenomenon where the dispersion can be observed in the vicinity of two local maxima of the density distribution. In particular, it uses a stochastic differential equation that leads to the limiting density of the volatility to be equal to a convex combination of two gamma densities, which has been proposed in \cite{iscam05}. However, the results are derived for a general process (\ref{slowly-osc}), using the limiting distribution and its statistical moments.} in 
\cite[Lemma 1]{stehlikova-sevcovic-kybernetika}. The differentials of the Wiener processes $d w_y$ and $d w_r$ are assumed to be uncorrelated.

The paper provides a tool for modeling the effects of rapidly oscillating stochastic volatility that can be observed in real markets, cf. \cite{sircar1},  \cite{sircar2}. If the length of the time scale for dispersion $y$ is denoted by $\varepsilon$, the equation 
(\ref{slowly-osc}) for the variable $y$ reads as follows:
\begin{equation}
d y=\frac{\alpha(y)}{\varepsilon}\, d t + \frac{v \sqrt{y}}{\sqrt{\varepsilon}} \, d w_y.
\label{rapid-osc}
\end{equation}
In what follows we will assume that $0< \varepsilon \ll 1$ is a small singular parameter.
The density of the conditional distribution of the process is given by the solution to Fokker-Planck equation. 
The density $g(y)$ of its stationary distribution, which is widely used in the computations  from \cite{stehlikova-sevcovic-kybernetika},  is then given by the normalized solution to the stationary Fokker-Planck equation, which reads as
\begin{equation}
\frac{v^2}{2} \partial^2_{y}(yg) - \partial_y(\alpha(y)g) =0  \label{eq:stacionarnyFP}
\end{equation}
for the process (\ref{rapid-osc}). 
Notice that the limiting density function $g$   is independent of the scaling parameter $\varepsilon>0$.

The market prices of risk functions are considered to be in the form 
$\tilde\lambda_1(t,r,y)=\lambda_1 \sqrt{r}\sqrt{y}$, $ \tilde\lambda_2(t,r,y)=\lambda_2 \sqrt{y}$, 
where $\lambda_1,\lambda_2\in \mathbb{R}$ are constants (note that this is a generalization of the original one-factor CIR model which assumes 
the market price of risk to be proportional to the square root of the short rate $r$). Then, we 
 rewrite the partial differential equation for the bond price $P$ in the operator form:
\begin{equation}
(\varepsilon^{-1} \mathcal{L}_0 + \varepsilon^{-1/2} \mathcal{L}_1 + \mathcal{L}_2 ) P^{\varepsilon} = 0,
\label{eq:bondprice2}
\end{equation}
where the linear differential operators $\mathcal{L}_0, \mathcal{L}_1,  \mathcal{L}_2$ are defined as follows:
$$
\mathcal{L}_0=\alpha(y) \partial_{y} + \frac {1}{2} v^2 y \partial^2_{y}, \:
\mathcal{L}_1=   - \lambda_2 v y \partial_{y},  \:
\mathcal{L}_2= \partial_{t} + (\kappa (\theta-r) - \lambda_1 r y ) \partial_{r} + \frac{1}{2} r y \partial^2_{rr} - r. 
$$
Next we expand the solution $P^{\varepsilon}$ into Taylor power series:
\begin{equation}
\label{expansion}
P^{\varepsilon}(t,r,y)= \sum_{j=0}^\infty \varepsilon^{\frac{j}{2}}  P_j(t,r,y)
\end{equation}
with the terminal conditions $P_0(T,r,y)=1, P_j(T,r,y)=0 \; \textrm{for } j \geq 1$ at expiry $t=T$.
The main result of this paper is the examination the singular limiting behavior of a solution $P^\varepsilon$ as $\varepsilon\to 0^+$. More precisely, it determines the first three terms $P_0,P_1,P_2$ of the asymptotic expansion (\ref{expansion}). 

The main tool in the derivation is averaging with respect to the limiting distribution, whose density $g$ is given by 
(\ref{eq:stacionarnyFP}), and is  denoted by brackets $\langle \cdot \rangle$  in the following. In particular, the following two propositions are essential: Firstly, a function $\psi$, for which $\mathcal{L}_0\psi$ is bounded, satisfies $\langle \mathcal{L}_0 \psi \rangle =0$ (see \cite[Lemma 3]{stehlikova-sevcovic-kybernetika}). Secondly, \cite[Lemma 4]{stehlikova-sevcovic-kybernetika} gives an expression for $\psi_y$ and $\langle \mathcal{L}_1 \psi \rangle $, where $\psi$ is a solution of $\mathcal{L}_0 \psi = F$ with the right-hand side being a given function satisfying $\langle F \rangle =0$.

The solution $P^\varepsilon = P^\varepsilon(t,r,y)$ of the bond pricing equation (\ref{eq:bondprice2})  can be approximated, for small values of the singular parameter $0<\varepsilon\ll 1$, by $$P^\varepsilon(t,r,y) \approx P_0(t,r) +\sqrt{\varepsilon} P_1(t,r) + \varepsilon P_2(t,r,y) + O(\varepsilon^\frac32)$$ and the main result of the paper lies in the derivation of the functions $P_0,P_1,P_2$.  
Note that the first two terms $P_0,P_1$ are independent of the $y$-variable representing unobserved stochastic volatility. 

The first term $P_0$ is a solution to the averaged equation $\langle \mathcal{L}_2 \rangle P_0=0$, which is the partial differential equation for the bond price in one-factor CIR model with parameters set to the averaged values (with respect to the limiting distribution) from the model studied here. It has a form
\begin{equation}
P_0(t,r) = A_0(t) e^{-B(t) r}, \label{eq:p0-form}
\end{equation}
where the functions $A_0$ and $B$ are given by a system of ordinary differential equations which can be solved in a closed form.
Neither the second term $P_1$ depends on the instantaneous level of the process $y$. The equation for the $P_1$ reads as
$$\langle \mathcal{L}_2 P_1 \rangle = f(t) r e^{-B(t)r},$$
where the function $B$ comes from (\ref{eq:p0-form}) and the function $f$ is obtained from the model parameters and the solution (\ref{eq:p0-form}) in a closed form. The solution has the form
$P_1(t,r) = (A_{10}(t) + A_{11}(t)r) e^{-B(t) r}$
with the function $B$ being the same as in (\ref{eq:p0-form}) and the functions $A_{10}, A_{11}$ satisfying  a system of linear ordinary differential equation.
The next term in the expansion, $P_2$, non-trivially depends on the $y$-variable.  It is decomposed into its expected value and zero-mean fluctuations as
$$P_2(t,r,y)=\bar{P}_2(t,r) + \tilde{P}_2(t,r,y)$$
where $\langle \tilde{P}_2 \rangle =0$. The function $\tilde{P}_2$ can be computed by integration, using the results obtained so far. The function $\bar{P}_2$ satisfies the equation  
$$\langle \mathcal{L}_2 \bar{P}_2 \rangle = (a(t)+b(t)r+c(t)r^2)e^{-B(t)r},$$
where the functions $a,b,c$ are given, and therefore has the form
$
\bar P_2(t,r) = (A_{20}(t) + A_{21}(t) r + A_{22}(t) r^2) e^{-B(t) r}$
where the function $B$ is the same as in (\ref{eq:p0-form}) and the functions $A_{20}, A_{21}, A_{22}$ are solutions to a linear system of ODEs. More detailed computations can be found in \cite{stehlikova-sevcovic-kybernetika}.

Recall the Fong-Vasicek model with stochastic volatility in which the short rate is given by the following pair of stochastic differential equation
\begin{eqnarray}
d r &=& \kappa_1 (\theta_1 - r) \: d t + \sqrt{y} \:  d w_1, \nonumber \\
d y &=& \kappa_2 (\theta_2 - y) \: d t +  v \sqrt{y} \:  d w_2. \label{eq:fvas}
\end{eqnarray}
For a suitable choices of market prices of risk, computation of the bond prices can be reduced into solving ordinary differential equations. This computational simplicity makes it a suitable choice for assessing the quality of the approximation of the kind described above. Introducing fast time scale of volatility, the equation (\ref{eq:fvas}) becomes (cf. equation (\ref{rapid-osc}))
\begin{equation}
d y = \frac{\kappa_2}{\varepsilon} (\theta_2 - y) \: d t +  \frac{v}{\sqrt{\varepsilon}} \sqrt{y} \:  d w_2. \label{eq:fvas2}
\end{equation}
However, when estimating parameters using the real data,  from the parameters of (\ref{eq:fvas2}) we are able to obtain only $\theta_2$, $\tilde{\kappa}_2= \frac{\kappa_2}{\varepsilon}$ and $\tilde{v}=\frac{\kappa_2}{\sqrt{\varepsilon}}$. Hence, we are not able to reconstruct three parameters $\kappa_2, v, \varepsilon$ from two values
$\tilde{\kappa}_2, \tilde{v}$.

Therefore, in the master thesis by Sele\v c\'eniov\'a \cite{dp-seleceniova}, supervised by Stehl\'ikov\'a, another approach was used, following the parameterization used by Danilov and Mandal in \cite{danilov-mandal-1} and \cite{danilov-mandal-2}. Strong mean-reversion in the process for volatility can be characterized by a large value of $\kappa_y$. Hence we can define $\varepsilon=1/\kappa_y$ and expect it to be small enough to be used as a perturbation parameter. In \cite{dp-seleceniova}, the derivation similar to that above has been made to compute the first two terms of the bond price expansion, leading to the  approximation of the bond price  of order zero 
$$P^\varepsilon(t,r,y) \approx P_0(t,r)$$ and of order one $$P^\varepsilon(t,r,y) \approx P_0(t,r) +\sqrt{\varepsilon} P_1(t,r).$$ Then, the resulting interest rates were compared with exact values. In Table \ref{tab:seleceniova} we present sample results. 

\begin{table}
\caption{{Interest rates from Fong-Vasicek model: comparison of order 0 and order 1 approximations with the exact values. Parameters are taken to be equal to $\kappa_1=0.109, \kappa_2 = 1.482, \theta_1 = 0.0652, \theta_2 - 0.000264, v= 0.01934, \lambda_1 = -11, \lambda_2 = -6, r=0.04$.  Source: Sele\v c\'eniov\'a, \cite{dp-seleceniova}.}}
\centering
\small
\smallskip
\begin{tabular}{c|ccc|cc}
& & exact interest rate & & \multicolumn{2}{c}{approximation} \\
maturity & $y=1.6 \times 10^{-4}$  &  $y=2.4 \times 10^{-4}$    
&  $y=3.2 \times 10^{-4}$ & order 0  & order 1\\
\hline 
1 & 0.0424  &  0.0426  & 0.0429 & 0.0427 & 0.0432 \\
2  & 0.0448  & 0.0451  & 0.0455 & 0.0451 & 0.0458 \\
3  & 0.0470  & 0.0474  & 0.0478 & 0.0473 & 0.0482 \\
4  & 0.0491  & 0.0495  & 0.0498 & 0.0493 & 0.0502 \\
5  & 0.0510  & 0.0514  & 0.0517 & 0.0511 & 0.0521 \\
6  & 0.0527  & 0.0531  & 0.0534 & 0.0528 & 0.0538 \\
7  & 0.0543  & 0.0547  & 0.0550 & 0.0543 & 0.0553 \\
8  & 0.0558  & 0.0561  & 0.0564 & 0.0557 & 0.0567 \\
9  & 0.0572  & 0.0575  & 0.0578 & 0.0570 & 0.0580 \\
10 & 0.0584  & 0.0587  & 0.0590 & 0.0582 & 0.0592 \\
\hline
\end{tabular}

\label{tab:seleceniova} 
\end{table}

Let us remark that even though the zero-order approximation of the bond
price equals to the bond price from one-factor model with averaged
coefficients, this is not the averaged bond price $\langle P(t,r,y) \rangle$. There is
even a stronger result: The averaged bond price $\langle P(t,r,y) \rangle$, although it is
a function of $t$ and $r$, does not equal to the bond price in any one-factor
model, as it has been shown in \cite{miyazaki} which is also reprinted at the end of this thesis.

\subsection{Convergence multiple-factor models}

The idea of approximating the bond prices in a model with general volatility by substituting the instantaneous volatility into a simple model of Vasicek type (i.e., with constant volatility) has been successfully applied also in multi-factor models:

Convergence models form a special class of two-factor models. A convergence model is used to model the entry of observed country into the monetary union (EMU). It describes the behavior of two short\--term interest rates, the domestic one and the instantaneous short rate for EMU countries. European short rate is modeled using a one-factor model. It is assumed to have an influence on the evolution of the domestic short rate and hence it enters the SDE for its evolution. This kind of model was proposed for the first time in \cite{corzo-schwarz}. The model is based on Vasicek model, the volatilities of the short rates are constant. Analogical model of Cox-Ingersoll-Ross type, where the volatilities are proportional to the square root of the short rate, was considered in \cite{lacko-dp} and \cite{lacko-dphannover}. In the following sections we describe these two models and show how they price the bonds. Then we present a generalization with nonlinear volatility, which is analogous to the volatility in one-factor CKLS model.

Let us consider a model defined by the following system of SDEs:
\begin{eqnarray}\label{eq:2FactorModel}
d r &=& \mu _r  (r,x,t)d t + \sigma_r (r,x,t)d w_1, \nonumber \\
d x &=& \mu _x  (r,x,t)d t + \sigma_x (r,x,t)d w_2,
\end{eqnarray}
where $\rho \in (-1,1)$ is the correlation between the increments of Wiener processes $W_1$ and $W_2$, i.e. $Cov(dW_1,dW_2)=\rho\,  d t$. Process $x$ is a random process, which is connected with instantaneous rate. It can be a long-term interest rate, a short-term interest rate in another country, etc. Relations between real and risk\--neutral parameters are analogous as in the one-factor case:
\begin{eqnarray*}
(\text{risk\--neutral drift function})_r = (\text{real drift function})_r - \lambda_r(r,x,t)\times(\text{volatility})_r, \\
(\text{risk\--neutral drift function})_x = (\text{real drift function})_x - \lambda_x(r,x,t)\times(\text{volatility})_x,
\end{eqnarray*}
where $\lambda_r$, $\lambda_x$ are market prices of risk of the short rate and the factor $x$ respectively.

If the short rate satisfies SDE (\ref{eq:2FactorModel}) in the real measure and market prices of risk are $\lambda_r(r,x,t), \lambda_x(r,x,t)$, then the bond price $P$ satisfies the following PDE (assuming that the factor $x$ is positive):
\begin{eqnarray*}
\frac{\partial P}{\partial t}+(\mu_r(r,x,t)-\lambda_r(r,x,t)\sigma_r(r,x,t))\frac{\partial P}{\partial r}+(\mu_x(r,x,t)-\lambda_x(r,x,t)\sigma_x(r,x,t))\frac{\partial P}{\partial x}\\
+\frac{\sigma_r(r,x,t)^2}{2}\frac{\partial^2 P}{\partial r}
+ \frac{\sigma_x(r,x,t)^2}{2}\frac{\partial^2 P}{\partial x}+\rho \sigma_r(r,x,t) \sigma_x(r,x,t) \frac{\partial^2 P}{\partial r \partial x}-rP=0\\
\end{eqnarray*}
for $r, x > 0$, $t \in (0,T)$
and the terminal condition $P(r,x,T)=1$ for $r, x > 0$. The PDE is derived using It$\hat{\text{o}}$ lemma and construction of risk-less portfolio, see, e.g. \cite{kwok},\cite{brigo-mercurio}.

\subsubsection{Convergence model of the CKLS type}

The paper \cite{zikova-stehlikova} is focused a convergence model of the CKLS type. Recall that the exact bond prices are known in the case of Vasicek-type model and their computation can be simplified to numerical solution of ordinary differential equations in the case of CIR-type model with uncorrelated increments of the two Wiener processes. In \cite{zikova-stehlikova}, the general CKLS model with uncorrelated Wiener processes (the effect of correlation can be seen only in higher order terms, when taking $\tau$ as a small parameter, numerical results presented in the paper show that the difference often occurs on decimal places which are not observable taking the precision of market quotes into account) is considered. Approximation formula from \cite{stehlikova} described in the previous section is used to compute European bond prices and in an analogous way, an approximation for domestic bond prices is proposed. It is tested numerically for CIR-type model and a general order of accuracy is derived. Then, a calibration procedure is suggested, tested on simulated data and applied to read data. The simple form of the approximation again allows relatively simple calibration procedure. 

\subsubsection{A three-factor convergence model}

A one-factor model is not always sufficient to model the European short rate in convergence model (as suggested by calibration results in \cite{zikova-stehlikova}), which affects also the appropriateness of the convergence model for the domestic currency. In paper \cite{stehlikova-zikova} by Stehl\'ikov\'a and Z\'ikov\'a, a three factor convergence model is suggested and provides first steps in the  analysis of approximation formulae for domestic bond prices. The European short rate is modeled as a sum of two CKLS-type factor, as described in the previous point, and the domestic rate follows a process reverting to the European rate. 

The fit of the convergence model  from \cite{zikova-stehlikova} suggests looking for a more suitable approximation of the short rate. The  paper \cite{halgasova-stehlikova-zikova} by Halga\v sov\'a, Stehl\'ikov\'a and Z\'ikov\'a studies an  estimation the short rate together with parameters of the model in Vasicek model. It is based on noting that for Vasicek model, the objective function (\ref{ucelova-funkcia})  for the calibration  is quadratic not only in parameters $\alpha$ and $\sigma^2$, but also in values of the short rates $r_1,\dots,r_n$.

\begin{figure}
 \includegraphics[width=0.48\textwidth]{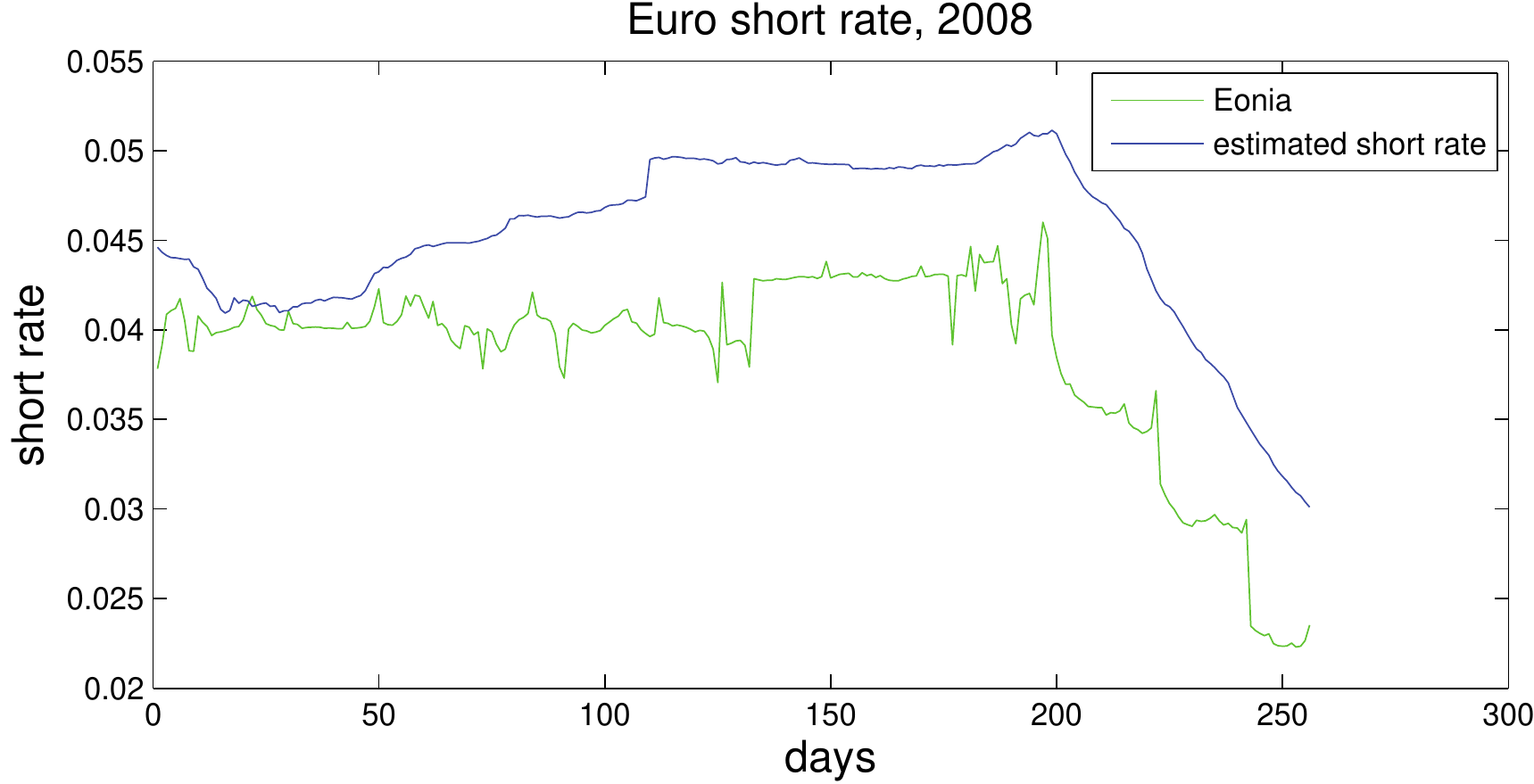}  
\ \ 
 \includegraphics[width=0.48\textwidth]{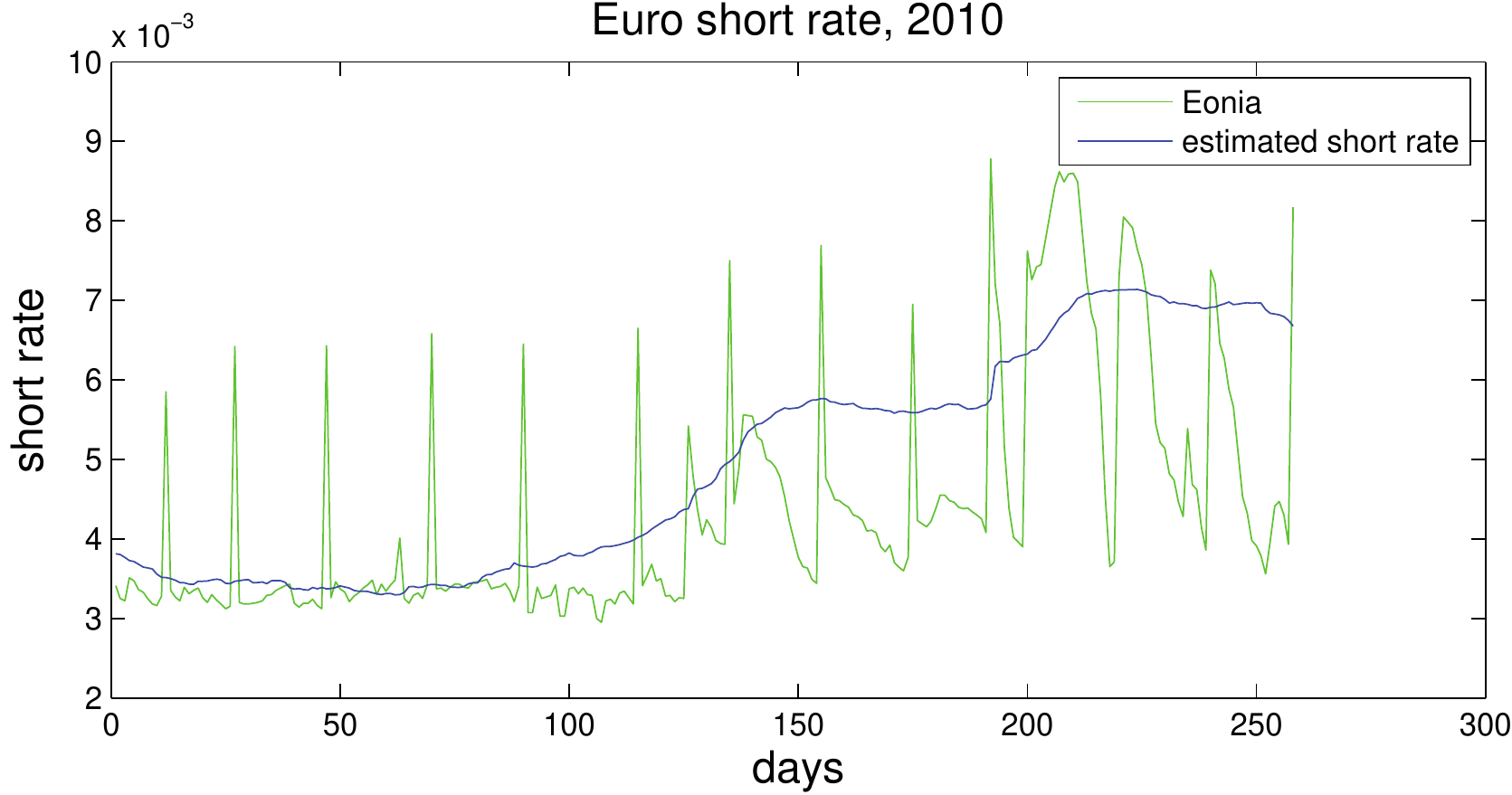} 
\caption{Estimating the short rate from Euribor term structures and its comparison with overnight rate Eonia. Source: Halga\v sov\'a, Stehl\'ikov\'a, Z\'ikov\'a, \cite{halgasova-stehlikova-zikova}.}
\label{fig:halgasova-stehlikova-zikova}
\end{figure}

Figure \ref{fig:halgasova-stehlikova-zikova} shows a comparison of the estimated short rate from Euribor term structures with a market overnight rate. The choice of the time frame for the calibration was motivated by a possible use as an input for a convergence model: Slovakia adopted the Euro currency in 2009 and Estonia in 2011.

Using the approximation of the bond prices in the CKLS model, this algorithm can be modified for estimating the short rate also in the CKLS model. This has been done in the master thesis \cite{dp-mosny} by Mosn\'y, supervised by Stehl\'ikov\'a. In the case of a general CKLS model, the objective function is not quadratic, but it is proposed to make a substitution
$y_i = \sigma^2 r_i^{2 \gamma}$
in the objective function, which results in the new objective function
which we minimize with respect to $\alpha, \beta, \sigma^2$ (model parameters), $r_1, \dots, r_n$ (short rates), $y_1, \dots, y_n$ (auxiliary variables treated as independent in the first step). In this way, for each $\beta$  a quadratic optimization problem is solved. For each $\beta$, there is therefore the optimal value of $\tilde{F}$ which is then used to find the optimal value of $\beta$.
Note that the variables $r_i$ and $y_i$ are not independent, the ratio $y_i/r_i^{2 \gamma}$ is equal to $\sigma^2$. 
By treating them as independent variables,  $y_i$ can be seen as approximations of $\sigma^2 r_i^{2 \gamma}$ when using real data. 
Hence the ratios $y_i/r_i^{2 \gamma}$  should provide a good approximation to $\sigma^2$. It is estimated as a median of these ratios.

\subsubsection{Convergence model of Vasicek type}
The first convergence model was proposed in the paper \cite{corzo-schwarz} by Corzo and Schwartz in the real probability measure:
\begin{eqnarray}\label{eq:VasicekModel1}
d r_d &=& \left(a+b\left(r_e-r_d\right)\right)d t + \sigma_d d w_d, 
\nonumber 
\\
d r_e &=& \left(c\left(d-r_e\right)\right)d t + \sigma_e d w_e,
\end{eqnarray}
where $Cov(dW_1,dW_2)=\rho d t$.
They considered constant market prices of risk, i.~e. $\lambda_d(r_d,r_e,\tau)=\lambda_d$ and $\lambda_e(r_d,r_e,\tau)=\lambda_e$. Hence for the European interest rate we have one-factor Vasicek model and we can easily price European bonds. Coefficient $b > 0$ expresses the power of attracting the domestic short rate to the European one with the possibility of deviation determined by the coefficient $a$. Rewriting the model into risk\--neutral measure we obtain:
\begin{eqnarray}\label{eq:VasicekModel2}
d r_d &=& \left(a+b\left(r_e-r_d\right)- \lambda_d \sigma_d\right)d t + \sigma_d d w_d, \nonumber 
\\
d r_e &=& \left(c\left(d-r_e\right) - \lambda_e \sigma_e\right)d t + \sigma_e d w_e,
\end{eqnarray}
where $Cov[dW_d,dW_e] = \rho d t$. We consider a more general model in risk\--neutral measure, in which the risk\--neutral drift of the domestic short rate is given by a general linear function of variables $r_d$, $r_e$ and the risk\--neutral drift of the European short rate is a general linear function of $r_e$. It means that the evolution of the domestic and the European short rates is given by:
\begin{eqnarray}\label{eq:VasicekModel3}
d r_d &=& \left(a_1+a_2r_d+a_3r_e\right)d t + \sigma_d d w_d, \\
d r_e &=& \left(b_1+b_2r_e\right)d t + \sigma_e d w_e,
\end{eqnarray}
where $Cov[dW_d,dW_e] = \rho d t$. Note that the system (\ref{eq:VasicekModel3}) corresponds to the system (\ref{eq:VasicekModel2}) with $a_1=a-\lambda_d\sigma_d$, $a_2=-b$, $a_3=b$, $b_1=cd-\lambda_e\sigma_e$, $b_2=-c$. Price $P(r_d,r_e,\tau)$ of a bond with time to maturity $\tau=T-t$ then satisfies the PDE:
\begin{eqnarray}\label{eq:PDRvas}
-\frac{\partial P}{\partial \tau}+(a_1+a_2r_d+a_3r_e)\frac{\partial P}{\partial r_d}+(b_1+b_2r_e)\frac{\partial P}{\partial r_e}\nonumber 
\\
+\frac{\sigma_d^2}{2}\frac{\partial^2 P}{\partial r_d^2}+\frac{\sigma_e^2}{2}\frac{\partial^2 P}{\partial r_e^2}+\rho \sigma_d \sigma_e \frac{\partial^2P}{\partial r_d \partial r_e} -r_dP&=0,
\end{eqnarray}
for $r_d,r_e>0,$ $\tau \in (0,T)$ and the initial condition $P(r_d,r_e,0)=1$ for $r_d, r_e > 0.$
Its solution can be found in the same way as in the original paper \cite{corzo-schwarz}. Assuming the solution in the form
\begin{eqnarray}\label{eq:SolutionInSeparateForm}
P(r_d,r_e,\tau)=e^{A(\tau)-D(\tau)r_d-U(\tau)r_e}, 
\end{eqnarray}
and setting it into the equation (\ref{eq:PDRvas}) we obtain the system of ordinary differential equations (ODEs):
\begin{eqnarray}\label{eq:EquationsVas}
\dot{D}(\tau)&=&1+a_2D(\tau),\nonumber \\
\dot{U}(\tau)&=&a_3D(\tau)+b_2U(\tau),\\
\dot{A}(\tau)&=&-a_1D(\tau)-b_1U(\tau)+\frac{\sigma_d^2 D^2(\tau)}{2}+\frac{\sigma_e^2 U^2(\tau)}{2}+\rho \sigma_d \sigma_e D(\tau) U(\tau)\nonumber 
\end{eqnarray}
with initial conditions $A(0)=D(0)=U(0)=0$. The solution of this system is given by:
\begin{eqnarray}\label{eq:SOLUTIONvas2}
D(\tau)&=&\frac{-1+e^{a_2\tau}}{a_2},\nonumber \\
U(\tau)&=& \frac{a_3\bigl(a_2-a_2 e^{b_2\tau}+b_2 
\left( -1+e^{a_2\tau}\right) \bigl)}{a_2\left(a_2-b_2\right)b_2},\\
A(\tau)&=&\int_0^{\tau} -a_1D(s)-b_1U(s)+\frac{\sigma_d^2 D^2(s)}{2}+\frac{\sigma_e^2 U^2(s)}{2}+\rho \sigma_d \sigma_e D(s) U(s) \mbox{d}s.\nonumber 
\end{eqnarray}
Note that the function $A(\tau)$ can be easily written in the closed form without an integral. We leave it in this form for the sake of brevity. Furthermore, we consider only the case when $a_2 \neq b_2$. If $a_2 = b_2$, then $U(\tau)$ has another form, but it is a very special case and we will not consider it further.

\subsubsection{Convergence model of CIR type}
Firstly we formulate the convergence model of CIR type (i.e. the volatilities are proportional to the square root of the short rates) in the real measure. 
\begin{eqnarray}\label{eq:CIRmodelREAL}
d r_d &=& \left(a+b\left(r_e-r_d\right)\right)d t + \sigma_d \sqrt{r_d}d w_d, \nonumber\\
d r_e &=& \left(c\left(d-r_e\right)\right)d t + \sigma_e \sqrt{r_e}d w_e,
\end{eqnarray}
where $Cov[dW_d,dW_e] = \rho d t$. If we assume the market prices of risk equal to $\lambda_e\sqrt{r_e}$, $\lambda_d\sqrt{r_d}$ we obtain risk neutral processes of the form:
\begin{eqnarray}\label{eq:CIRmodel1}
d r_d &=& \left(a_1+a_2r_d+a_3r_e\right)d t + \sigma_d \sqrt{r_d} d w_d, \nonumber \\
d r_e &=& \left(b_1+b_2r_e\right)d t + \sigma_e \sqrt{r_e}d w_e, 
\end{eqnarray}
where $Cov[dW_d,dW_e] = \rho d t$. In what follows we consider this general risk\-- neutral formulation (\ref{eq:CIRmodel1}).

The European short rate is described by one-factor CIR model, so we are able to price European bonds using an explicit formula. 
Price of domestic bond $P(r_d,r_e,\tau)$ with maturity $\tau$ satisfies the PDE
\begin{eqnarray}\label{eq:PDRcir}
-\frac{\partial P}{\partial \tau}+(a_1+a_2r_d+a_3r_e)\frac{\partial P}{\partial r_d}+(b_1+b_2r_e)\frac{\partial P}{\partial r_e} \nonumber \\
+\frac{\sigma_d^2r_d^{2}}{2}\frac{\partial^2 P}{\partial r_d^2}+\frac{\sigma_e^2r_e^{2}}{2}\frac{\partial^2 P}{\partial r_e^2}+\rho \sigma_d \sqrt{r_d} \sigma_e \sqrt{r_e} \frac{\partial^2P}{\partial r_d \partial r_e} -r_dP&=0,
\end{eqnarray}
for $r_d,r_e>0, \tau \in (0,T)$
with the initial condition $P(r_d,r_e,0)=1$ for $r_d, r_e > 0.$
It was shown in \cite{lacko-dp} (in a slightly different parametrization of the model) that solution in the form (\ref{eq:SolutionInSeparateForm}) exists only when $\rho=0$. In this case we obtain system of ODEs
\begin{eqnarray}\label{eq:EquationsCIR}
\dot{D}(\tau)&=&1+a_2D(\tau)-\frac{\sigma_d^2 D^2(\tau)}{2},\nonumber \\
\dot{U}(\tau)&=&a_3D(\tau)+b_2U(\tau)-\frac{\sigma_e^2 U^2(\tau)}{2},\\
\dot{A}(\tau)&=&-a_1D(\tau)-b_1U(\tau),\nonumber
\end{eqnarray}
with initial conditions $A(0)= D(0) = U(0) = 0,$ which can be solved numerically. 

\subsubsection{Convergence model of CKLS type}
We consider a model in which risk\--neutral drift of the European short rate $r_e$ is a linear function of $r_e$, risk\--neutral drift of the domestic short rate $r_d$ is a linear function of $r_d$ and $r_e$ and volatilities take the form $\sigma_er_e^{\gamma_e}$ and $\sigma_dr_d^{\gamma_d}$, i.e.

\begin{eqnarray}\label{eq:CKLSmodel}
d r_d &=& (a_1+a_2 r_d + a_3 r_e)d t + \sigma_d r_d^{\gamma_d} dw_d, \nonumber \\
d r_e &=& (b_1+b_2r_e)d t + \sigma_e r_e^{\gamma_e} dw_e,
\end{eqnarray}
where $Cov[dW_d,dW_e] = \rho d t$. Parameters $a_1, a_2, a_3, b_1, b_2 \in \mathbb{R}, \sigma_d, \sigma_e > 0, \gamma_d, \gamma_e \geq 0$ are given constants and $\rho \in (-1,1)$ is a constant correlation between the increments of Wiener processes $d W_d$ a $d W_e$. We will refer to this model as \emph{two-factor convergence model of Chan\--Karolyi\--Longstaff\--Sanders (CKLS) type}.
The domestic bond price $P(r_d, r_e, \tau)$ with the maturity $\tau$ satisfies PDE:
\begin{eqnarray}\label{eq:PDRckls}
-\frac{\partial P}{\partial \tau}&+&(a_1+a_2r_d+a_3r_e)\frac{\partial P}{\partial r_d}+(b_1+b_2r_e)\frac{\partial P}{\partial r_e} \nonumber \\
&+&\frac{\sigma_d^2r_d^{2\gamma_d}}{2}\frac{\partial^2 P}{\partial r_d^2}+\frac{\sigma_e^2r_e^{2\gamma_e}}{2}\frac{\partial^2 P}{\partial r_e^2}+\rho \sigma_d r_d^{\gamma_d} \sigma_e r_e^{\gamma_e} \frac{\partial^2P}{\partial r_d \partial r_e} -r_dP=0,
\end{eqnarray}
for $r_d,r_e>0, \tau \in (0,T),$ with initial condition $P(r_d,r_e,0)=1$ for $r_d, r_e > 0.$
Unlike for Vasicek and uncorrelated CIR model, in this case it is not possible to find solution in the separable form (\ref{eq:SolutionInSeparateForm}).
For this reason, we are looking for an approximative solution.

\subsection{Approximation of the domestic bond price solution}
The bond prices in the CKLS type convergence model are not known in a closed form. This is already the case for the European bonds, i.e. one-factor CKLS model. We use the approximation from \cite{stehlikova}. In this approximation we consider one-factor Vasicek model with the same risk\--neutral drift and we set current volatility $\sigma r^{\gamma}$ instead of constant volatility into the closed form formula for the bond prices. We obtain
\begin{equation}\label{eq:Apr1factorCKLS}
\ln P^{ap}_{e}(\tau,r)=\left(\frac{b_1}{b_2}+\frac{\sigma^2r^{2\gamma}}{2b_2^2} \right) \left(\frac{1-e^{b_2\tau}}{b_2}+\tau \right)+\frac{\sigma^2r^{2\gamma}}{4b_2^3}\left(1-e^{b_2 \tau}\right)^2 + \frac{1-e^{b_2\tau}}{b_2}r.
\end{equation} 
We use this approach to propose an approximation for the domestic bond prices. We consider the domestic bond prices in Vasicek convergence model with the same risk\--neutral drift and we set $\sigma_dr_d^{\gamma_d}$ instead of $\sigma_d$ and $\sigma_er_e^{\gamma_e}$ instead of $\sigma_e$ into (\ref{eq:SOLUTIONvas2}). Hence, we have
\begin{equation}\label{eq:Apr2factorCKLS}
\ln P^{ap}=A-Dr_d-Ur_e
\end{equation}
\vspace{-0.2cm}
where
\begin{eqnarray*}
D(\tau)&=&\frac{-1+e^{a_2\tau}}{a_2},\nonumber \\
U(\tau)&=& \frac{a_3\bigl(a_2-a_2 e^{b_2\tau}+b_2 
\left( -1+e^{a_2\tau}\right) \bigl)}{a_2\left(a_2-b_2\right)b_2},\nonumber \\
A(\tau)&=&\int_0^{\tau} -a_1D(s)-b_1U(s)+\frac{\sigma_d^2r_d^{2\gamma_d} D^2(s)}{2}+\frac{\sigma_e^2r_e^{2\gamma_e} U^2(s)}{2}\\
&+&\rho \sigma_dr_d^{\gamma_d} \sigma_er_e^{\gamma_e} D(s) U(s) \mbox{d}s.\nonumber 
\end{eqnarray*}

In the CIR convergence model the domestic  bond price $P^{CIR, \rho=0}$ has a separable form (\ref{eq:SolutionInSeparateForm}) and functions $A, D, U$ are characterized by a system of ODEs (\ref{eq:EquationsCIR}). This enables us to compute Taylor expansion of its logarithm around $\tau=0$. We can compare it with the expansion of proposed approximation $\ln P^{CIR, \rho=0, ap}$ (computed either using its closed form expression (\ref{eq:Apr2factorCKLS}) or the system of ODEs (\ref{eq:SOLUTIONvas2}) for Vasicek convergence model). More detailed computation can be found in \cite{ZIKOVA}. In this way we obtain the accuracy of the approximation for the CIR model with zero correlation: 
\begin{equation}\label{eq:AccuracyCIR}
\text{ln}P^{CIR,\rho=0,ap}-\text{ln}P^{CIR,\rho=0}=\frac{1}{24}\left(-a_2\sigma_d^2r_d-a_1\sigma_d^2-a_3\sigma_d^2r_e \right)\tau^4 + o(\tau^4)\\
\end{equation}
for $\tau \rightarrow 0^+$.

Let us consider real measure parameters: $a=0$, $b=2$, $\sigma_d=0.03$, $c=0.2$, $d=0.01$, $\sigma_e=0.01$ and market price of risk $\lambda_d=-0.25$, $\lambda_e=-0.1$. In the risk\--neutral setting (\ref{eq:CIRmodel1}) we have $a_1=a-\lambda_d\sigma_d=0.0075$, $a_2=-b=-2$, $a_3=b=2$, $b_1=cd-\lambda_e\sigma_e=0.003$, $b_2=-c=-0.2$, $\sigma_d=0.03$, $\sigma_e=0.01$. With the initial values for the short rates $r_d = 1.7 \%$ a $r_e = 1 \%$ we generate the evolution of domestic and European short rates using Euler\--Maruyama discretization.  In Table \ref{tab:rozdielCIR} we compare the exact interest rate and the approximative interest rate given by (\ref{eq:Apr2factorCKLS}). We observe very small differences. Note that the Euribor market data are quoted with the accuracy $10^{-3}$. Choosing other days, with other combination of $r_d$, $r_e$, leads to very similar results. The difference between exact and approximative interest rate remains nearly the same.

\begin{table}
\begin{tabular}{c c}
\begin{minipage}[t]{.48\linewidth}
\centering
\begin{center}
\footnotesize{
\begin{tabular}{c|c|c|c}
\textbf{Mat.} & \textbf{Exact} & \textbf{Approx.} & \textbf{Diff.} \\
 \textbf{[year]} & \textbf{yield [\%]} & \textbf{yield [\%]} & \textbf{[\%]} \\
\hline 
 $1/4$ &  1.63257 & 1.63256  & 7.1 E-006  \\
 $1/2$ &  1.58685 & 1.58684 & 1.4 E-005 \\
 $3/4$ & 1.55614  & 1.55614 & 4.8 E-006  \\
 $1$ &  1.53593 &  1.53592  & 1.1 E-005  \\
 $5$ & 1.56154  & 1.56155 &  -5.0 E-006 \\
 $10$ & 1.65315  & 1.65323  &  -8.3 E-005 \\
 $20$ & 1.74696  & 1.74722 &  -2.5 E-004 \\
 $30$ &  1.78751 & 1.78787 & -3.7 E-004  \\
\hline
\end{tabular}}
\end{center}
\end{minipage}
\begin{minipage}[t]{.48\linewidth}
\centering
\begin{center}
\footnotesize{
\begin{tabular}{c|c|c|c}
\textbf{Mat.} & \textbf{Exact} & \textbf{Approx.} & \textbf{Diff.} \\
 \textbf{[year]} & \textbf{yield [\%]} & \textbf{yield [\%]} & \textbf{[\%]} \\
\hline 
$1/4$ &  1.08249 &  1.08250  & -8.2 E-006  \\
$1/2$ & 1.15994  & 1.15996 & -1.7 E-005  \\
$3/4$ & 1.21963  & 1.21964 & -7.0 E-006  \\
$1$ & 1.26669  &  1.26671  &  -1.6 E-005 \\
$5$ &  1.53685 & 1.53691 &  -6.2 E-005 \\
$10$ & 1.65113   & 1.65127  & -1.4 E-004  \\
$20$ & 1.74855  &  1.74884  &  -2.9 E-004 \\
$30$ & 1.78879  & 1.78918 &  -3.9 E-004 \\
\hline
\end{tabular}}
\end{center}
\end{minipage}
\end{tabular}
\caption{\footnotesize{Exact and approximative domestic yield for 1st (left) observed day, $r_d = 1.7 \%$, $r_e = 1 \%$ and for 252nd (right) observed day, $r_d = 1.75 \%$, $r_e = 1.06 \%$}.}\label{tab:rozdielCIR}
\end{table}

Finally, we present a detailed derivation of  the order of accuracy of the proposed approximation in the general case. We use analogous method as in \cite{stehlikova} and \cite{stehlikova-sevcovic} for one-factor models 
and in \cite{lacko-dp} to study the influence of correlation $\rho$ on bond prices in the convergence CIR model.

Let $f^{ex}=\ln P^{ex}$ be the logarithm of the exact price $P^{ex}$ of the domestic bond in two factor convergence model of CKLS type. It satisfies the PDE (\ref{eq:PDRckls}). Let $f^{ap}=\ln P^{ap}$ be the logarithm of the approximative price $P^{ap}$ for the domestic bond price given by (\ref{eq:Apr2factorCKLS}).
By setting $f^{ap}$ to the left-hand side of (\ref{eq:PDRckls}) we obtain non-zero right-hand side, which we denote as $h(r_d, r_e, \tau)$. We expand it into Taylor expansion and obtain that 
\begin{equation}\label{h}
h(r_d,r_e,\tau)=k_3(r_d,r_e)\tau ^3 + k_4(r_d,r_e)\tau ^4 + o(\tau ^4),
\end{equation}
for $\tau \rightarrow 0^+$, where
\begin{equation*}\label{k3}
k_3(r_d,r_e)=\frac{1}{6}\sigma_d^{2}\gamma_d r_d^{2\gamma_d-2}\left( 2a_1r_d + 2a_2r_d^2 + 2a_3r_dr_e - r_d^{2\gamma_d}\sigma_d^2 + 2\gamma_dr_d^{2\gamma_d}\sigma_d^2 \right),
\end{equation*}
\begin{eqnarray*}\label{k4}
k_4(r_d,r_e)&=& \frac{1}{48} \frac{1}{r_e^2} r_d^{-2 +\gamma_d} \sigma_d 
\Bigr(12 a_2^2 \gamma_d r_d^{2 + \gamma_d} r_e^2 \sigma_d - 16 \gamma_d r_d^{1 + 3 \gamma_d} r_e^2 \sigma_d^3 + 
6 a_3 b_1 \gamma_e r_d^2 r_e^{1 + \gamma_e} \rho \sigma_e \\
&+&6 a_3 b_2 \gamma_e r_d^2 r_e^{2 + \gamma_e} \rho \sigma_e + 
6 a_3^2 \gamma_d r_d r_e^{3 + \gamma_e} \rho \sigma_e- 3 a_3 \gamma_d r_d^{2 \gamma_d} r_e^{2 + \gamma_e} \rho \sigma_d^2 \sigma_e\\
&+& 3 a_3\gamma_d^2 r_d^{2\gamma_d}r_e^{2+\gamma_e}\rho\sigma_d^2\sigma_e +6 a_3\gamma_d\gamma_e r_d^{1+\gamma_d} r_e^{1+2 \gamma_e}\rho^2 \sigma_d \sigma_e^2- 
3 a_3 \gamma_e r_d^2 r_e^{3 \gamma_e} \rho \sigma_e^3\\
&+& 3 a_3 \gamma_e^2 r_d^2 r_e^{3 \gamma_e} \rho \sigma_e^3 + 
6 a_1 \gamma_d r_d r_e^2 \left(2 a_2 r_d^{\gamma_d} \sigma_d + a_3 r_e^{\gamma_e} \rho \sigma_e\right)\\ 
&+& 6 a_2 \gamma_d r_e^2 \bigl(\left(-1 + 2 \gamma_d\right) r_d^{3 \gamma_d} \sigma_d^3 + 
a_3 r_d \left(2 r_d^{\gamma_d} r_e \sigma_d + r_d r_e^{\gamma_e} \rho \sigma_e\right)\bigl)\Bigr).
\end{eqnarray*}

We define function $g(\tau, r_d, r_e):=f^{ap}-f^{ex}=\ln P^{ap}-\ln P^{ex}$ as a difference between logarithm of the approximation and the exact price.
Using the PDEs satisfied by $f^{ex}$ and $f^{ap}$ we obtain the following  PDE for the function $g$:
{\small
\begin{eqnarray}\label{PDRg}
-\frac{\partial g}{\partial \tau}&+&\left(a_1+a_2r_d+a_3r_e\right)\frac{\partial g}{\partial r_d}+\left(b_1+b_2r_e\right)\frac{\partial g}{\partial r_e} +
\frac{\sigma_d^2 r_d^{2\gamma_d}}{2} \left[\left( \frac{\partial g}{\partial r_d}\right)^2 + \frac{\partial^2 g}{\partial r_d^2}\right] \nonumber \\
&+& \frac{\sigma_e^2 r_e^{2\gamma_e}}{2} \left[\left( \frac{\partial g}{\partial r_e}\right)^2 + \frac{\partial^2 g}{\partial r_d^2}\right]
+\rho \sigma_d r_d^{\gamma_d} \sigma_e r_e^{\gamma_e} \left(\frac{\partial g}{\partial r_d} \frac{\partial g}{\partial r_e} +
\frac{\partial ^2 g}{\partial r_d \partial r_e}  \right) \\
= h(r_d,r_e,\tau) &+& \frac{\sigma_d^2 r_d^{2\gamma_d}}{2} \left[ \left(\frac{\partial f^{ex}}{\partial r_d}\right)^2 - 
\frac{\partial f^{ap}}{\partial r_d}\frac{\partial f^{ex}}{\partial r_d} \right]
+ \frac{\sigma_e^2 r_e^{2\gamma_e}}{2} \left[ \left(\frac{\partial f^{ex}}{\partial r_e}\right)^2 - 
\frac{\partial f^{ap}}{\partial r_e}\frac{\partial f^{ex}}{\partial r_e} \right]\nonumber \\
&+& \rho \sigma_d r_d^{\gamma_d} \sigma_e r_e^{\gamma_e} \left[ 2 \frac{\partial f^{ex}}{\partial r_d}\frac{\partial f^{ex}}{\partial r_e}
-\frac{\partial f^{ap}}{\partial r_d}\frac{\partial f^{ex}}{\partial r_e}
-\frac{\partial f^{ex}}{\partial r_d}\frac{\partial f^{ap}}{\partial r_e} \right].\nonumber
\end{eqnarray}
}

Suppose that $g(r_d,r_e,\tau)=\sum_{k=\omega}^{\infty}c_{k}(r_d,r_e)\tau^{k}$. 
For $\tau=0$ is both the exact and approximative bond price equal to one, so  $f^{ex}(r_d,r_e,0)=f^{ap}(r_d,r_e,0)=0$. It means that $\omega>0$ and on the left hand side of the equation (\ref{PDRg}) the term with the lowest order is $c_{\omega}\omega\tau^{\omega-1}$. Now we investigate the order of the right hand side of the equation.

We know that  $f^{ex}(r_d,r_e,0)=0$. It means that $f^{ex}=O(\tau)$ and also partial derivation $\frac{\partial f^{ex}}{\partial r_d}$ and $\frac{\partial f^{ex}}{\partial r_e}$ are of the order $O(\tau)$. From the approximation formula (\ref{eq:Apr2factorCKLS}) we can see that $\frac{\partial f^{ap}}{\partial r_d}=O(\tau)$, 
$\frac{\partial f^{ap}}{\partial r_e}=O(\tau^2)$.
Since $h(r_d, r_e, \tau)=O(\tau^3)$, the right hand side of the equation (\ref{PDRg}) is at least of the order $\tau^2$. The left hand side of the equation (\ref{PDRg}) is of the order $\tau^{\omega-1}$ and hence $\omega-1 \geq 2$, i.e. $\omega \geq 3$.
It means that
$$f^{ap}(r_d,r_e,\tau)-f^{ex}(r_d,r_e,\tau)=O(\tau^3).$$
Using this expression we can improve estimation of the derivative $\frac{\partial f^{ex}}{\partial r_e}$ as follows:
$\frac{\partial f^{ex}}{\partial r_e}=\frac{\partial f^{ap}}{\partial r_e}+O(\tau^3)=O(\tau^2)+O(\tau^3)=O(\tau^2).$
We also estimate the terms on the right hand side in the equation (\ref{PDRg}):
\begin{eqnarray}\label{odhad1}
\left(\frac{\partial f^{ex}}{\partial r_d}\right)^2 - 
\frac{\partial f^{ap}}{\partial r_d}\frac{\partial f^{ex}}{\partial r_d} &=& \frac{\partial f^{ex}}{\partial r_d}
\left(\frac{\partial f^{ex}}{\partial r_d}-\frac{\partial f^{ap}}{\partial r_d}\right)=O(\tau).O(\tau^3)=O(\tau^4),
\end{eqnarray}

\begin{eqnarray}\label{odhad1.5}
\left(\frac{\partial f^{ex}}{\partial r_e}\right)^2 - 
\frac{\partial f^{ap}}{\partial r_e}\frac{\partial f^{ex}}{\partial r_e}= \frac{\partial f^{ex}}{\partial r_e}
\left(\frac{\partial f^{ex}}{\partial r_e}-\frac{\partial f^{ap}}{\partial r_e}\right)=O(\tau^2).O(\tau^3)=O(\tau^5),
\end{eqnarray}

\begin{eqnarray}\label{odhad2}
2 \frac{\partial f^{ex}}{\partial r_d}\frac{\partial f^{ex}}{\partial r_e}
- \frac{\partial f^{ap}}{\partial r_d}\frac{\partial f^{ex}}{\partial r_e}
-\frac{\partial f^{ex}}{\partial r_d}\frac{\partial f^{ap}}{\partial r_e} = 
\frac{\partial f^{ex}}{\partial r_d}
\left(\frac{\partial f^{ex}}{\partial r_e}-\frac{\partial f^{ap}}{\partial r_e}\right)\nonumber \\
+ \frac{\partial f^{ex}}{\partial r_e}
\left(\frac{\partial f^{ex}}{\partial r_d}-\frac{\partial f^{ap}}{\partial r_d}\right) = O(\tau).O(\tau^3)+O(\tau^2).O(\tau^3)=O(\tau^4)+O(\tau^5)=O(\tau^4).
\end{eqnarray}
Since $h(r_d,r_e,\tau)=O(\tau^3)$, the right hand side of the equation (\ref{PDRg}) is $O(\tau^3)$ and the coefficient at $\tau^3$ is the coefficient of the function $h(r_d,r_e,\tau)$ at $\tau^3$, i.e. $k_3(r_d,r_e)$. It means that $\omega=4$ and comparing the coefficients at $\tau^3$ on the left and right-hand side of (\ref{PDRg}) we obtain
$-4c_4(r_d,r_e)=k_3(r_d,r_e),$
i.e.
$c_4(r_d,r_e)=-\frac{1}{4}k_3(r_d,r_e).$
Hence we have proved the following theorem. 
\begin{theorem}\label{VETApresnost}
Let $P^{ex}(r_d,r_e,\tau)$ be the price of the domestic bond in two-factor CKLS convergence model, i.e. satisfying equation (\ref{eq:PDRckls}) and let $P^{ap}$ be the approximative solution defined by (\ref{eq:Apr2factorCKLS}). Then 
$$\ln P^{ap}(r_d,r_e,\tau)-\ln P^{ex}(r_d,r_e,\tau)=c_4(r_d,r_e)\tau^4+ o(\tau^4)$$
for $\tau \rightarrow 0^+,$
where coefficient $c_4$ is given by 
\begin{equation}\label{c4}
c_4(r_d,r_e)=-\frac{1}{24}\sigma_d^{2}\gamma_d r_d^{2\gamma_d-2} 
\left( 2a_1r_d + 2a_2r_d^2 + 2a_3r_dr_e - r_d^{2\gamma_d}\sigma_d^2 + 2\gamma_dr_d^{2\gamma_d}\sigma_d^2 \right).
\end{equation}
\end{theorem}

Note that if we substitute  $\gamma_d=\frac{1}{2}$ and $\rho=0$ into Theorem \ref{VETApresnost}, we obtain the formula (\ref{eq:AccuracyCIR}) for CIR model derived earlier in (\ref{eq:AccuracyCIR}).

In some cases it is possible to improve an approximation by calculating more terms in Taylor expansion of the function $g=\ln P^{ap}-\ln P^{ex}$. It is so also in this case. 
Using that $f^{ap}-f^{ex}=O(\tau^4)$, we are able to improve estimates (\ref{odhad1}) and (\ref{odhad2})
and to deduce that also the coefficient at $\tau^4$ on the right hand side of equation (\ref{PDRg}) comes only from the function $h$.
Hence it is equal to $k_4(r_d,r_e)$, which is given by (\ref{k4}). Comparing coefficients at $\tau^4$ on the left and right hand side of (\ref{PDRg}) we obtain:
\begin{eqnarray*}
- 5c_5+(a_1+a_2r_d+a_3r_e)\frac{\partial c_4}{\partial r_d}+(b_1+b_2r_e)\frac{\partial c_4}{\partial r_e}\\
+\frac{\sigma_d^2r_d^{2\gamma_d}}{2}\frac{\partial^2 c_4}{\partial r_d^2}+\frac{\sigma_e^2r_e^{2\gamma_e}}{2}\frac{\partial^2 c_4}{\partial r_e^2}+4\rho\sigma_d r_d^{\gamma_d}\sigma_e r_e^{\gamma_e}\frac{\partial^2 c_4}{\partial r_d \partial r_e}=k_4,
\end{eqnarray*}
which enables us to express $c_5$ using already known quantities.

Let us define an approximation $\ln P^{ap2}$ by:
$$\ln P^{ap2}(r_d,r_e,\tau)=\ln P^{ap}-c_4(r_d,r_e)\tau^4-c_5(r_d,r_e)\tau^5.$$
Then 
$\ln P^{ap2}- \ln P^{ex}=O(\tau^6)$
and therefore the new approximation $\ln P^{ap2}$ is of the order $O(\tau^6).$

\subsection{Financial interpretation of the short rate factors and their evolution}

In the PhD thesis by \v Sest\'ak \cite{phd-sestak}, supervised by \v Sev\v covi\v c, the approximation formula from  \cite{dp-halgasova} is used to estimate the model for European countries. The rate for each country is decomposed into a risk-free rate (common to all the countries) and a credit spread (specific for each country). The formula from \cite{dp-halgasova} is used to price bonds in this setting. The author suggests a calibration procedure which is computationally demanding since it involves a large data set - yields of all countries considered simultaneously (it is not possible to split this for each country, since the risk-free rate, which is one of the outputs, is shared by all the countries). Hence a simple approximate formula for the bond prices is crucial for a successful estimation.

\begin{figure}

\begin{center}
\includegraphics[width=0.7\textwidth]{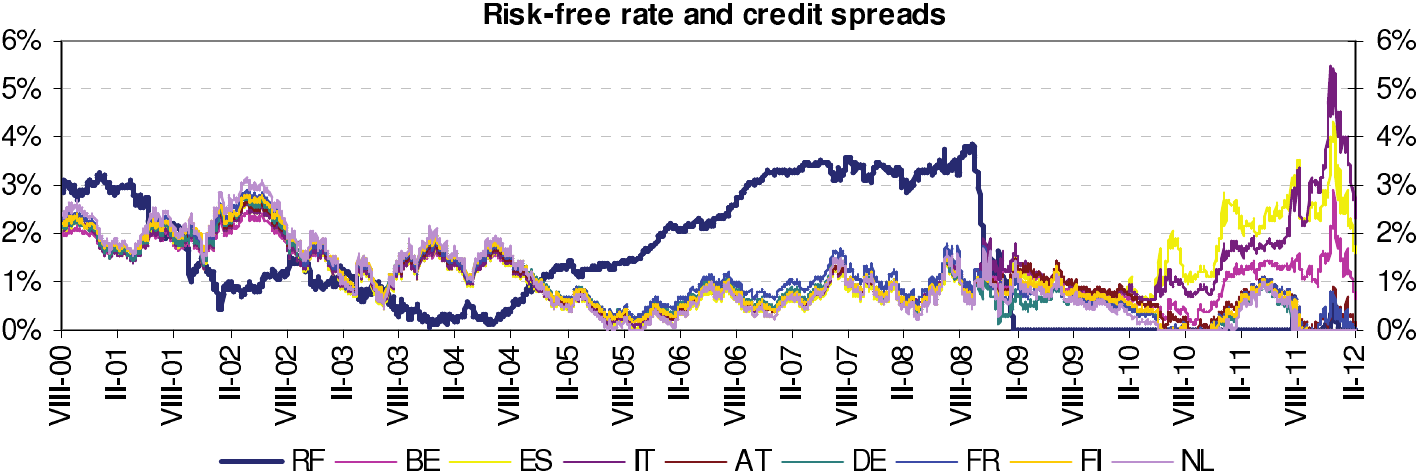} 
\\
\includegraphics[width=0.7\textwidth]{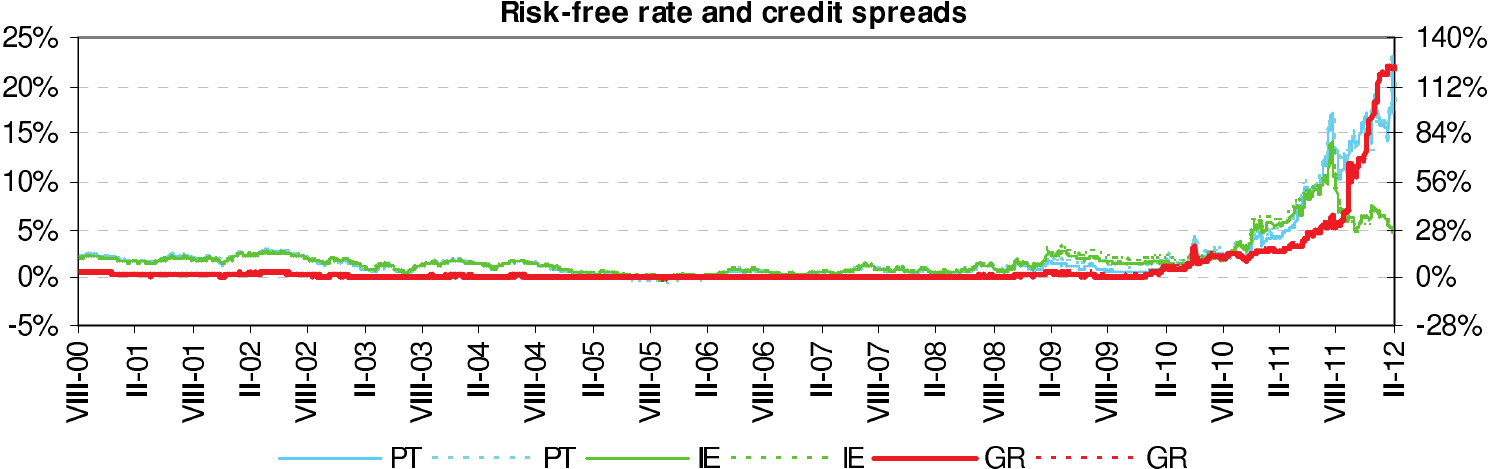}  
\end{center}

\caption{Estimating the risk-free rate and credit spread in the European countries. In the figure below, the values for Greece are shown in the right axis, for the other countries in the left axis. Source: \v Sest\'ak, \cite{phd-sestak}.}
\label{fig:sestak}
\end{figure}

Figure \ref{fig:sestak} shows results of the estimation from \cite{phd-sestak}. Note how the very different evolution of the credit spread for Greece starts from a certain time, compared to the values obtained for the other counties.

\section{Conclusions}

\fancyhead[EC,OC]{}
\fancyhead[EL,OR]{\thepage}

In this survey we presented an overview of short rate models and presented some of the approaches to compute approximations of bond prices where the exact solutions are not available.

Firstly, we considered one-factor models. The simple models of Vasicek and Cox-Ingersoll-Ross admit closed form bond prices and therefore can serve as either basis for construction of analytical approximations or as testing cases for assessing numerical accuracy of different approximation formulae. Using partial differential approach to bond pricing enables us to derive their order of accuracy for small times remaining to maturity.

In the second part we dealt with multi-factor models – the process for the short rate written as a sum of  two factors, second factor being the stochastic volatility or the European interest rate when modeling rates in a country before adoption of the Euro currency. In case of convergence model we provided also an example of a three-factor model, in which the European rate is modeled by a two-factor model. We studied similar analytic approximations for convergence models as in the case of one-factors models. Here we provided also a proof of accuracy of the proposed approximation; similar reasoning was applied also in the analyses of other models, where we only stated the results. Moreover, we studied the asymptotics of a fast time scale of volatility in stochastic volatility models.


\end{document}